\documentclass{article}

\PassOptionsToPackage{numbers, compress}{natbib}

 \usepackage[main, final]{neurips_2026}

\usepackage{amsmath}
\usepackage[utf8]{inputenc} %
\usepackage[T1]{fontenc}    %
\usepackage[pagebackref=true,breaklinks=true,letterpaper=true,colorlinks,bookmarks=false,citecolor=CornflowerBlue,linkcolor=Bittersweet]{hyperref}      %
\usepackage{url}            %
\usepackage{booktabs}       %
\usepackage{amsfonts}       %
\usepackage{nicefrac}       %
\usepackage{microtype}      %
\usepackage{xcolor}         %
\usepackage{graphicx}
\usepackage[most]{tcolorbox}
 \usepackage{enumitem}
 \usepackage{algorithm}
 \usepackage{algorithmic}
 \usepackage{amssymb}
\usepackage{sidecap}
\usepackage{caption}
\usepackage{multirow}
\definecolor{eqcolor}{HTML}{6699CC}
\usepackage{wrapfig}
\usepackage{subcaption}
\usepackage{tipa}

\newtcolorbox{mybox}[1][]{
    ams align,
    colback=white,
    colframe=eqcolor,
    #1
}

\title{Reconstructing the Vocal Tract with Differentiable Acoustic Simulation}

\author{%
  Eric M. Chen \\
  MIT CSAIL \\ 
  \texttt{echen01@mit.edu} \\
  \And
  Jin Woo Lee \\
  MIT RLE \&
  KAIST GSCT \\
  \texttt{jnlee@mit.edu} \\
 \And
  Vincent Sitzmann \\
    MIT CSAIL \\
  \texttt{sitzmann@mit.edu} 
}

\usepackage[dvipsnames]{xcolor}

\newif\ifdraft
\drafttrue
\ifdraft
\newcommand{\smc}[1]{{\color{blue}[\textbf{SM:} #1]}}

\else
\newcommand{\smc}[1]{}

\fi

\newcommand{\re}{\text{Re}}

\begin{document}

\maketitle

\begin{abstract}
The vocal tract is the region of the human body responsible for filtering one's voice to create speech. In this paper, we present a differentiable and GPU accelerated acoustic simulator for the vocal tract. The differentiable simulator synthesizes speech by propagating sound along an acoustic tube model of the vocal tract, and via its gradients, can solve the inverse problem: reconstructing the shape of the vocal tract solely from the sound it produces. Although the inverse mapping between geometry and sound is notoriously non-convex, we discover that gradient descent succeeds with three technical contributions: (1) we design a frequency domain formulation of the vocal tract's fluid dynamics that is 70x more GPU parallelizable than finite differences in time, (2) we integrate a differentiable model for turbulence to synthesize consonants, and (3) similar to prior work in implicit neural representations (INRs) and neural fields, we find that parameterizing the geometry with a neural network accelerates convergence and escapes local minima that trap discrete representations. Because the simulator is differentiable, it is readily integrated with other deep learning pipelines to enable novel linguistics and medical imaging applications. (1) We demonstrate self-supervised autoencoding of vocal tract shapes across 11 languages, and (2) we couple our simulator with a generative model of MRI (magnetic resonance imaging) images to reconstruct one's moving vocal tract from only their speech without paired data.

\end{abstract}

\section{Introduction}
Learning to speak is fundamentally an inverse problem. An infant babbles to discover the non-linear mapping between their articulatory motor commands and their speech, learning the controls required to produce language. Most remarkably, children solve this inverse problem without ever seeing paired data between the muscle configuration of their vocal tract and the speech it produces. 

In domains like robotics, this type of sensorimotor learning is accelerated by differentiable simulators. Differentiable simulators, such as Brax~\citep{brax2021github}, Taichi~\citep{Hu2019DiffTaichiDP}, and Warp~\citep{warp2022}, are computer graphics engines which allow gradients from a physical environment to backpropagate directly to a policy, enabling efficient learning of locomotion and manipulation. But to picture the physical actions of your mouth, tongue, etc. that form language, no equivalent computational infrastructure exists. 

Such a method to visualize one's vocal tract would have significant impact in language, such as studying the cognitive processes of language acquisition~\citep{beguvs2023articulation,Caren2024}, in music, as a tool for voice coaching or singing synthesis, as well as in healthcare such as speech pathology and speech brain-computer interfaces~\citep{Metzger2023AHN}. Although several differentiable acoustic simulators~\citep{wang2024hearing, finnendahl2025differentiable, lan2024acoustic,liang2023av} have been recently proposed for room acoustics, these rely on ray-based methods for stationary systems that are ill-suited for capturing the time-varying morphological transformations of the vocal tract~\citep{kuttruff2016room,glassner1989introduction}. 
On the other hand, while neural vocoders~\citep{oord2016wavenet, kong2020hifigan,wang2019neural} achieve high acoustic realism, they operate as black boxes that ignore the underlying physics. Conversely, classical vocal tract simulators~\citep{kelly_lochbaum_1962,Maeda1982} are computationally expensive and CPU-bound, rendering them incompatible with modern deep learning pipelines and making them ineffective at solving the speech-to-motor inverse problem. 

We address this gap by introducing a differentiable, GPU-accelerated physical simulator of speech production. Our simulator produces speech by propagating sound along an acoustic tube representation of a person's vocal tract, and via its gradients, can visualize the shape of the vocal tract from only audio. We adopt a frequency-domain solution tailored for GPUs. This approach is  $70 \times$ more parallelizable on GPUs and, crucially, avoids high-frequency noise inherent to sequential time-stepping (Section~\ref{subsec:fds}). And to support end-to-end learning, we introduce a gradient-friendly formulation of turbulent noise for modeling consonants (Section~\ref{subsec:consonants}). Although acoustic inverse problems are non-convex, leading to sub-optimal solutions~\citep{sudholt2023vocal,Panchapagesan2011Chain}, we find that, much like in neural rendering, representing the geometry with a neural network acts as a regularizer that stabilizes the learning dynamics (Section~\ref{subsec:neural_fields}). We validate these design choices empirically.

\begin{figure*}[t]
    \centering
    \includegraphics[width=\linewidth]{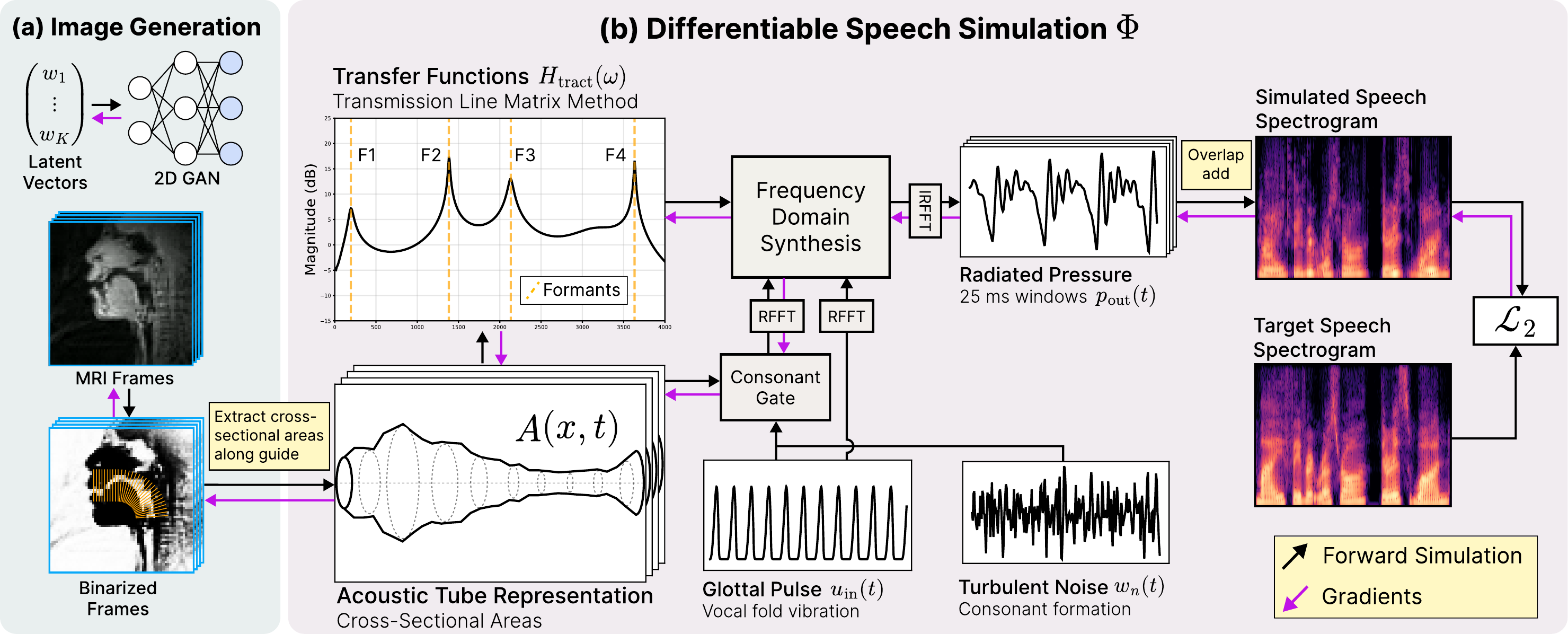}
    \caption{\textbf{Method Overview.} We introduce a method to reconstruct a MRI video of a person's moving vocal tract from their speech alone. \textbf{(a)} Given an MRI image of a person's vocal tract, we extract cross-sectional areas along the length of the tract. \textbf{(b)} These areas define an \textit{acoustic tube}. Our differentiable simulator $\Phi$ propagates sound through the acoustic tube to synthesize speech. Gradients from the loss flow back through $\Phi$ and the generative model into a set of latent vectors, reconstructing the MRI video from the target speech signal.}
    \label{fig:recon}
    \vspace{-10pt}
\end{figure*}

We demonstrate the effectiveness of our simulator through two novel applications. First, we use our simulator to train a \textit{self-supervised} autoencoder which maps raw audio to the vocal tract shape (Section~\ref{subsec:autoencoding}).
Second, we provide a tool to visualize an MRI (magnetic resonance imaging) video of a person's moving vocal tract from their speech, without training on any paired speech-MRI data. This is a task, to the best of our knowledge, that has not been previously attempted. As outlined in Figure~\ref{fig:recon}, we connect a generative model of MRI images to our differentiable acoustic simulator. Our method then jointly recovers an MRI video and audio. Because our model is geometrically grounded, it outperforms the closest prior work trained on paired data~\cite{Nguyen2024Speech2rtMRISD}.

All in all, by establishing a first-of-its-kind differentiable link between vocal tract geometry and speech, we provide the infrastructure to treat speech as a physically-grounded learning problem.

\section{Related Work}

\textbf{Articulatory Synthesis.}
Many physics-based simulators for the vocal tract have been introduced over the last century~\citep{kelly_lochbaum_1962,Maeda1982,Sondhi1987,BirkholzVTL}, the most well-known being the Kelly-Lochbaum method~\citep{kelly_lochbaum_1962}, and the state of the art being VocalTractLab (VTL)~\citep{BirkholzVTL}. However, these models are difficult to directly backpropagate through because they rely on finite differences to model a differential equation of sound propagation. Furthermore, they use separate forward models for consonants and vowels, causing a discontinuity in their control flow.

\textbf{Black Box Methods.}
To avoid backpropagating through an acoustic simulation, recent works have proposed to solve the inverse problem by training neural networks on paired data between speech and articulatory measurements. However, because these networks are black boxes, they are difficult to interpret physically. Furthermore, their quality fundamentally depends on paired training data, which are sparse. TensorTract2~\citep{Krug2025TT2} trains a neural network to map articulatory controls to synthetic sounds generated from VocalTractLab~\citep{BirkholzVTL}. Black box models have also been trained on biomedical data such as vocal tract MRI~\citep{Nguyen2024Speech2rtMRISD} and electromagnetic articulography (EMA)~\citep{Cho2024EMA}. In contrast, our method does not require paired articulatory and speech data. 

Another related line of work is differentiable digital signal processing (DDSP)~\citep{engelddsp}. DDSP models been used to model human speech, such as in neural source filter~\citep{wang2019neural} and ~\citet{SchulzeForster2022LSFDDSP}. However, unlike our method, these models are not constrained by physics.~\citet{SchulzeForster2022LSFDDSP}, for instance, model the vocal tract filter with an all-pole filter via line spectral frequencies. On the other hand, our physical simulator can model both poles and zeros, allowing for realistic modeling of radiation and viscous losses.  

\textbf{Differentiable Acoustic Simulation with Neural Fields.}
Following the success of differentiable rendering and neural fields for solving inverse problems in 3D computer vision~\citep{mildenhall2020nerf,sitzmann2020implicit}, several works have also proposed differentiable simulators and neural fields for sound propagation~\citep{luo2022learning,wang2024hearing, lan2024acoustic,liang2023av}. These works focus primarily on room-scale environments, where sound waves are approximated as rays for efficiency. However, such ray-based approximations are inappropriate for the human voice. This is because the fundamental pitch of the human voice (80--255 Hz) has much longer wavelength than the diameter of the vocal tract ($<10$ cm), so the wave effects become dominant. Instead, our simulator explicitly leverages physics-based modeling suitable for time-varying acoustic tubes such as the human vocal tract.

\citet{sudholt2023vocal} have also recently proposed to reconstruct vocal tract areas by backpropagating through a differentiable simulator. However, instead of parameterizing the vocal tract's area with an implicit neural representation, they adopt an explicit parameterization. Furthermore, they only demonstrate results on the limited setting of static vowels and consonants. Meanwhile, our work shows that neural fields are an effective way to reconstruct both the geometry and movement of the vocal tract, enabling the synthesis of full words.

\section{Background: The Acoustics of Speech}
\label{sec:background}
Speech production is commonly framed as a source-filter process. First, a source signal is produced by either vibrations of the vocal folds or turbulent noise formed at narrow constrictions. Then, the vocal tract's geometry acts as an acoustic resonator, filtering the audio source to shape the different phonemes we interpret as speech.

\textbf{Acoustic Tubes.} Although the space of all possible speech sounds is large, the vocal tract's geometry moves in relatively fewer dimensions over time. Following many prior works~\citep{BilbaoNumerical2009,kelly_lochbaum_1962,Maeda1982,BirkholzVTL}, we represent the vocal tract as an \textit{acoustic tube}: a one-dimensional tube whose cross-sectional area $A(x)$ varies along its length. Pictured in Figure~\ref{fig:recon}, the size of this tube is dynamically controlled by articulators such as the tongue, mouth and jaw. Sound propagates as waves along this tube, whose physics are governed by the linearized Euler equations:
\begin{align}
    \frac{A(x)}{\rho c^2} \frac{\partial}{\partial t}p(x, t) = -\frac{\partial}{\partial x}u(x, t)  & \label{eq:lee1} &   \frac{\rho}{A(x)} \frac{\partial}{\partial t}u(x, t) = -\frac{\partial}{\partial x}p(x, t) 
\end{align}
$p(x, t)$ is sound (pressure), $u(x, t)$ is the volume velocity of airflow, $\rho$ is the density of air, and $c$ is the speed of sound. $x$ ranges over $[0, L]$, where $L$ is the length of the vocal tract, and $t$ ranges over $[0, T]$, where $T$ is the length of the speech signal. Most importantly, $p(L, t)$, which we name $p_{\text{out}}(t)$, {is what we perceive as speech. }For brevity, material terms are omitted, but the equations can be augmented with  viscous damping and wall vibration terms, summarized in Appendix~\ref{app:webster_derivation}. To synthesize speech, boundary conditions are set at each end of the tract as Eq.~\eqref{eq:bc1}-\eqref{eq:bc2}. At $x=0$, the vocal folds vibrate, leading to a {glottal pulse}: $u(0, t) \triangleq u_\text{in}(t)$. We represent $u_\text{in}(t)$ with the Liljencrants--Fant (LF) model~\citep{fant1985four}. At $x=L$, a radiation boundary is set for the lips.

\textbf{Acoustic Tubes as Source-Filter Models.}
In this problem, the primary object of interest is determining the lip radiation pressure $p_{\mathrm{out}}(t)$ given a source $u_{\mathrm{in}}(t)$. Considering a sufficiently small time window compared to the changes in oral cavity, $A(x)$ can be regarded as quasi-static, and the relationship between $p_{\mathrm{out}}$ and $u_{\text{in}}$ is time-invariant. Solving the linearized Euler equations for a given $A(x)$ is thus analogous to deriving a linear time-invariant (LTI) filter $h_{\text{tract}}$, such that $p_{\text{out}} = h_{\text{tract}} * u_{\text{in}}$, or equivalently $P_{\text{out}} = H_{\text{tract}} U_{\text{in}}$ in the frequency domain. 
The peaks of $H_{\text{tract}}$'s frequency response function are called its formants, which are crucial auditory cues for humans to perceive vowels (illustrated in Figure~\ref{fig:recon}). Although representing the system in either the time domain or frequency domain may seem equivalent, from an optimization perspective, the choice of domain significantly impacts the convergence of gradient-based approaches as we shall see in Section~\ref{sec:dss}.

\textbf{Discretization.} In accordance with the conventions of prior studies \cite{kelly_lochbaum_1962,deller1993discrete}, the area function $A(x,t)$ of the dynamic oral cavity is simplified into a quasi-static and piecewise-constant geometry. Specifically, the area function $A(x, t)$ of a vocal tract is discretized into uniform $N\times K$ lattice with spatial sections of length $L/N$ and temporal windows of length $T/K$, where each windowed section is considered to have constant area, \textit{i.e.}, $A[n, k] = A\left( \frac{nL}{N}, \frac{kT}{K}\right)\approx\mathrm{const.}$ We call the simulator which solves the Euler equations in the time-varying case $\Phi: (u_{\text{in}}, A)\mapsto p_{\text{out}}$.
\begin{figure}[t]
\centering
\includegraphics[width=\linewidth]{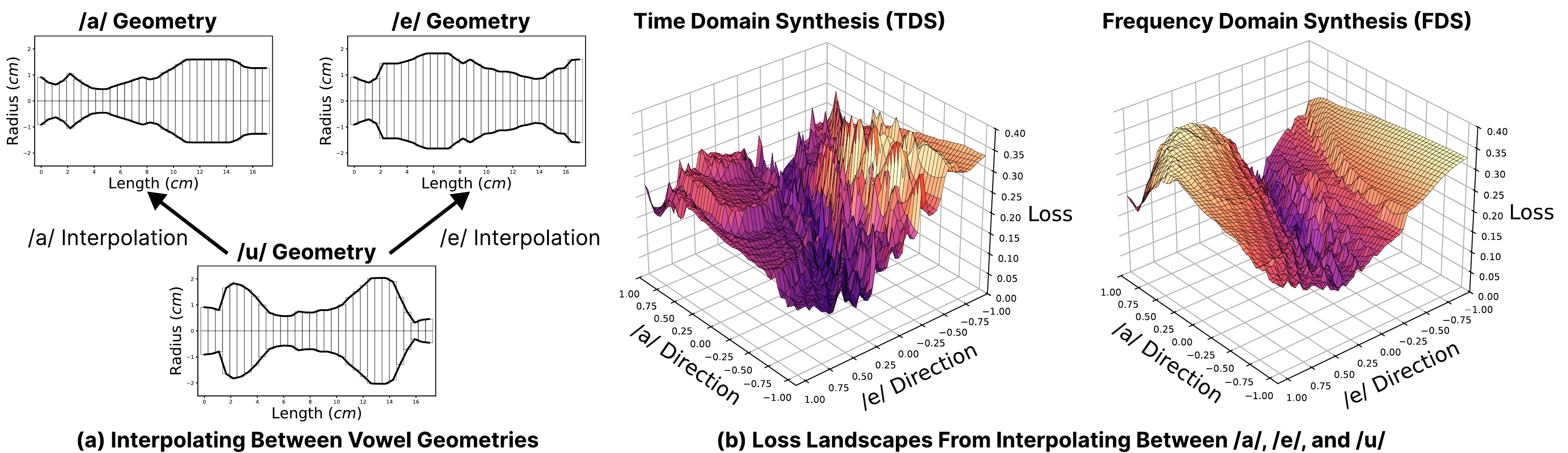}
\caption{\textbf{FDS Leads to Smoother Loss Landscapes.} \textbf{(a)} Starting from the vowel /u/, we interpolate its area in the direction of the /a/ geometry and /e/ geometry. \textbf{(b)} We plot the log Mel spectrogram loss between the sound produced by the interpolated geometries and the sound produced by /u/. The loss from time domain synthesis is noiser than the loss from frequency domain synthesis.}
\label{fig:noise}
\end{figure}

\label{subsec:background}
\section{Problem Statement and Method Overview} 
\label{sec:problem_statement}
 We outline our method in Figure~\ref{fig:recon}. Given a target speech sample $p_{\text{tgt}}$, our objective is to infer the geometry of the vocal tract that produced $p_{\text{tgt}}$. We first estimate the vocal fold vibration $u_{\text{in}}(t)$ with a pitch tracker (Appendix~\ref{app:experimental_details}), and set it as the input boundary condition. Then, our strategy is to leverage gradients from a differentiable simulator $\Phi$ to find an area function $A_\theta$ that minimizes the squared distance, $\mathcal{L}_2$, between spectrograms of the output $p_{\text{out}} \triangleq \Phi(u_{\text{in}}, A_\theta)$ and the target $p_{\text{tgt}}$. Enabling both efficient forward simulation and gradient-based reconstruction requires addressing a series of challenges, each of which motivates a technical contribution of our work.

 \textbf{How do we make $\Phi$ and its gradients efficient to compute?} Section~\ref{subsec:fds} demonstrates how solving the governing equations in the frequency domain enables GPU parallelization which is $70\times$ faster than time domain synthesis, and more importantly, enables stable gradient flow.
 
\textbf{How do we model non-linear effects?} Turbulence, which forms consonants, is a non-linear effect not captured by the linearized Euler equations. Section~\ref{subsec:consonants} introduces a differentiable turbulence model that unifies vowel and consonant synthesis in a single forward pass.
 
 \textbf{How do we avoid sub-optimal solutions for $A_\theta$?} The speech-to-geometry inverse problem is non-convex, and a differentiable forward model alone does not guarantee a good solution. Drawing on recent work in neural fields, we find that parameterizing $A_\theta$ with a neural network accelerates convergence. Furthermore, coupling our simulator with neural networks unlocks new applications in linguistics and medical imaging, described in Chapter~\ref{sec:vocal_recon}.
\section{Differentiable Speech Simulation}
\label{sec:dss}
\subsection{Frequency Domain Synthesis (FDS)}
\label{subsec:fds}

Various numerical methods such as finite differences can be employed to obtain solutions of the linearized Euler equations.
Even though backpropagating the gradients through the finite differences is possible, we find that the feasibility of this gradient signal is limited by challenges such as noise accumulation from extensive temporal recursion, leading to noisy loss landscapes that cause neural network training to fail in many cases. 
Assuming the linear time invariance as described in Section \ref{sec:background}, the system admits separable eigensolutions $p(x, t) = P(x, \omega)e^{-j\omega t}$ and $u(x, t) = U(x, \omega)e^{-j\omega t}$, and solving the linearized Euler's equations essentially reduces to the Helmholtz eigenvalue problem:  %
 \begin{align}
    \frac{j\omega A(x) }{\rho c^2} P(x, \omega) = \frac{\partial}{\partial x} U(x, \omega) & & 
      \frac{j \omega \rho}{A(x)} U(x, \omega) = \frac{\partial}{\partial x} P(x, \omega)    %
\end{align}
The boundary conditions simplify to $U(0, \omega) = U_\text{in}(\omega)$, and $P(L, \omega) = Z_\text{lips}(\omega) U(L, \omega)$, where $Z_{\text{lips}}(\omega)$ is the impedance defined in Eq.~\eqref{eq:Zrad}.The time-independence achieved by transforming the governing equation into a spatial ODE offers distinct advantages by not only eliminating temporal recursion but also enabling independent analysis of the geometries for each frequency $\omega$.
When using central differences for the spatial derivatives, as $\frac{\partial}{\partial x} U(x, \omega)\mid_{n+\frac{1}{2}} = \frac{U_{n+1} - U_n}{\Delta x}$ and $\frac{\partial}{\partial x} P(x, \omega)\mid_{n+\frac{1}{2}}  = \frac{P_{n+1} - P_n}{\Delta x}$, the relationship between neighboring sections $n$ and $n+1$ can be expressed as a $2\times2$ chain matrix, $\mathbf{K}_n$. Multiplying successive $\mathbf{K}_n$ matrices over $N$ sections derives a closed-form solution for the vocal tract filter, ${H}_{\text{tract}}(\omega)$:
\begin{align}\label{eq:transfer-function}
\begin{bmatrix}P_{n+1}\\ U_{n+1} \end{bmatrix} = \mathbf{K}_n\begin{bmatrix}P_{n}\\ U_{n} \end{bmatrix}&, \qquad \mathbf{K}_{\text{tot}}(\omega) = \prod_{n=0}^{N} \mathbf{K}_n(\omega), \qquad
    \begin{bmatrix}P_{N+1} \\ U_{N+1}\end{bmatrix} = \underbrace{\begin{bmatrix}\mathbf{A} & \mathbf{B}\\ \mathbf{C} & \mathbf{D}\end{bmatrix}}_{\mathbf{K}_{\text{tot}}(\omega)} \begin{bmatrix}P_0 \\ U_0\end{bmatrix} \\
    H_{\text{tract}}(\omega) &= \frac{Z_{\text{lips}}(\omega)}{\mathbf{A}(\omega) - \mathbf{C}(\omega)Z_{\text{lips}}(\omega)} = \frac{P_\text{out}(\omega)}{U_\text{in}(\omega)}&
\end{align}
The output pressure spectrum is then $P_{\text{out}} = H_{\text{tract}}U_{\text{in}}$, and the synthesized speech $p_{\text{out}}$ can be recovered with an inverse Fourier transform. This technique is an instance of the \textit{transmission line matrix} (TLM) method~\citep{Birkholz2004InfluenceOT}, because its equations describe an equivalent acoustic circuit consisting of $\mathbf{K_n}$ components. We detail its derivation in Appendix~\ref{subsec:transfer_function}.
Two properties of this formulation are critical. First, the closed-form solution for $H_{\text{tract}}$ is independent of temporal discretization. Second, the $\mathbf{K}_n$ matrices can be solved simultaneously for each frequency, enabling parallelization. 

\textbf{Modeling Time-Varying Vocal Tracts.}
\label{subsec:time_varying}
To synthesize speech from time-varying vocal tracts, we compute $p_{\text{out}}$ independently for successive windows then combine them using the overlap-add method.
We apply a Tukey analysis window with $\alpha=0.25$ to $u_{\text{in}}$, and a Hann synthesis window to $p_{\text{out}}$.

\begin{figure}[t] 
    \begin{minipage}{0.68\linewidth}
        \small
        \captionof{table}{\textbf{Speech Reconstruction Metrics.} When reconstructing real speech examples from LibriTTS-R~\citep{Koizumi2023LibriTTSRAR}, the audio from frequency domain synthesis (FDS) is significantly more perceptible than the results from time domain synthesis (TDS). To balance fidelity and efficiency, we use a window of 25 ms and a hop ratio of $1/4$. FDS is also 71.5x faster than real time (RT) while TDS is only 1.08x faster.}
        \scalebox{0.86}{
        \begin{tabular}{l|ccccc}
        \toprule
        &\textbf{TDS} & \multicolumn{2}{c}{\textbf{FDS (25 ms window)}} & \multicolumn{2}{c}{\textbf{ FDS (50 ms window)}}  \\
        \cmidrule(lr){2-2} \cmidrule(lr){3-4} \cmidrule(lr){5-6} 
        &  ---  & \textbf{1/8 Hop} & \textbf{1/4 Hop} & \textbf{1/8 Hop} & \textbf{1/4 Hop} \\
        \midrule
        SI-SDR & 7.32 {\scriptsize $\pm$2.42} & 17.83 {\scriptsize $\pm$2.19} & 16.39 {\scriptsize $\pm$2.42} & 17.47 {\scriptsize $\pm$1.94} & 15.82 {\scriptsize $\pm$2.47}\\
         STOI & 0.77 {\scriptsize $\pm$0.03} & 0.93 {\scriptsize $\pm$0.02} & 0.93 {\scriptsize $\pm$0.02} & 0.93{\scriptsize $\pm$0.02} & 0.93 {\scriptsize $\pm$0.02} \\
         PESQ & 1.55 {\scriptsize $\pm$0.14} & 2.02 {\scriptsize $\pm$0.31} & 1.98 {\scriptsize $\pm$0.26}& 1.97 {\scriptsize $\pm$0.26} & 1.98 {\scriptsize $\pm$0.25} \\
        RT Factor &  1.08\texttt{x}  & 35.0\texttt{x} & 71.5\texttt{x} & 36.0\texttt{x} & 70.0\texttt{x} \\
        \bottomrule
        \end{tabular}
        }
        \label{tab:perceptibility}
    \end{minipage}
    \hfill
    \begin{minipage}{0.3\linewidth}
        \includegraphics[width=\linewidth]{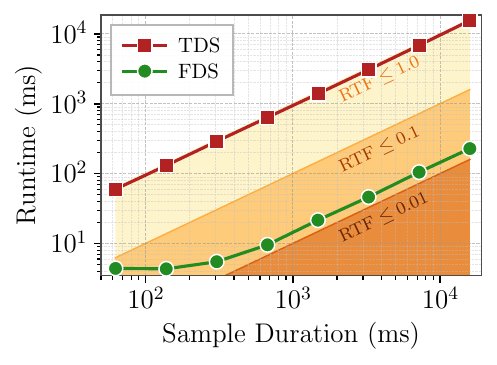}
        \caption{\textbf{Runtime.} FDS is 70x faster than TDS on GPU. Tested on a RTX 4090 with a 25 ms window and $1/4$ hop.}
        \label{fig:runtime}        
    \end{minipage}
\end{figure}

\subsection{Differentiable Gating of Consonants}
\label{subsec:consonants}
\textbf{Consonants as Turbulence.} Although the linearized Euler equations can be solved efficiently in the frequency domain, they do not describe consonant production, which is non-linear. Consonants are formed when steady air is pushed through narrow constrictions in the vocal tract, breaking down into turbulence. This transition from steady to turbulent flow is modeled by the squared Reynolds number. At section $A_n$ it is defined as
\begin{align}
    \re{}_n^2 = \frac{4\rho^2}{\pi \mu^2}\frac{u_n(t)^2}{A_n},
    \end{align}
where $\mu$ is the dynamic viscosity of air, and $u_n$ is the volume velocity at section $n$. $u_n$ is calculated in the frequency domain as $U_n = H_{0\mapsto n} U_{\text{in}}$~\footnote{The notation $H_{i\mapsto j}$ denotes the transfer function from section $i$ to $j$. Formulas are provided in Appendix~\ref{subsec:transfer_function}.}. Notice that as $A_n$ becomes smaller, such as the case when forming consonants by constricting one's teeth, tongue, or lips, the Reynolds number increases. 

\textbf{Discontinuities in Turbulence Models.} Since directly solving the compressible Navier--Stokes equations to capture turbulence is computationally prohibitive, a widely adopted approach is to use auxiliary models for the turbulence. However these auxiliary models have discontinuities that prevent gradient-based optimization. 
Prior works~\citep{Birkholz2006NoiseSA,Sondhi1987,Maeda1982} adopt consonant formation by injecting a noise source  into the vocal tract at the smallest point of constriction: $A_\text{min} = \min \{A_1, \dots, A_N\}$, with amplitude
    $u_{\text{noise}} = \max\{0, \alpha \cdot z_n  (\re{}^2_n - \re{}_{\text{crit}}^2)\}$. $\alpha$ is a gain, $z_n$ is noise, and $\re_{\text{crit}}$ is the critical Reynolds number. $\re_{\text{crit}}$ is a constant that gates the effect of turbulence. 

\textbf{Removing the Discontinuities.} We replace the $\min$ and $\max$ operations with smooth approximations. First, to differentiate through the position of the constriction, we simply replace the $\text{argmin}$ operator with $\text{softmax}(-A_n)$. By doing so, gradients flow backward into every area element, and turbulence is weighted by how small the area is at $A_n$. Second, we replace the $\max$ gate with a softplus function, so the overall noise injected at section $n$ is 
\begin{align}
    u_{\text{noise}_n} = \alpha \cdot z_n \cdot  \text{softmax}(-A_n) \cdot \text{softplus}(\re{}^2_n - \text{Re}_{\text{crit}}^2)
\end{align}
We use fractal Perlin noise~\citep{Perlin1985} as the noise source, which has been widely adopted to add turbulent textures in various fields such as computer graphics~\citep{Bridson2007CurlnoiseFP,Stam1993TurbulentWF} and sound synthesis~\citep{hahn1995integrated}.
$\re_{\text{crit}}$ is set to 3500 as experimentally determined by~\citet{Sondhi1987}.

\textbf{Unified Vowel and Consonant Synthesis.}
We represent the final form of our vocal tract transfer function as follows.
\begin{align}\label{eq:tf-tract-plus-noise}
    P_{\text{out}}(\omega) &= H_{\text{tract}}(\omega)U_{\text{in}}(\omega) + \sum_{n=1}^N H_{n\mapsto N+1}(\omega) U_{\text{noise}_n}(\omega)
\end{align}
This harmonic-plus-noise spectral modeling can also be viewed as encapsulating our vocal tract transfer function within a differentiable DSP pipeline \cite{engelddsp}, thereby not only enabling dynamic transition between vowels and consonants, but also preserving differentiability. Although turbulence is computed in the time domain, this technique is still 70x faster than real time (Figure~\ref{fig:runtime}).

\subsection{Comparing Frequency Domain Synthesis (FDS) to Time Domain Synthesis (TDS)}

\textbf{FDS Smooths Gradients.} To validate our choice of formulation, we compare frequency domain synthesis (FDS) against time domain synthesis (TDS) based on the methods of~\citep{Birkholz2004InfluenceOT}. The TDS method is a semi-implicit scheme described in Appendix~\ref{app:finite_differences}. While the two methods may be theoretically equivalent in continuous time, we find that TDS has significantly noiser gradients in implementation. To illustrate this, in Figure~\ref{fig:noise}, we visualize the loss landscapes of $\mathcal{L}_{2}$ when interpolating between three vowels. FDS produces a consistently smooth loss landscape, while TDS suffers from noise. 

\label{subsubsec:fds_eval}
\textbf{FDS Improves Acoustic Reconstruction.} The differences in loss landscape has direct consequences for optimization.  Table~\ref{tab:perceptibility} reports perceptual speech quality metrics for area functions fit to 25 speech samples from LibriTTS-R~\citep{Koizumi2023LibriTTSRAR}. The experimental setup is described in Appendix~\ref{app:experimental_details}.  Using TorchAudio-Squim~\citep{Kumar2023TorchaudioSquimRS}, we measure SI-SDR (scale-invariant signal-to-distortion ratio), STOI (short-time objective intelligibility), and PESQ (wideband perceptual evaluation of speech quality). Across all metrics, FDS substantially outperforms TDS. The SI-SDR for TDS is almost 10 dB lower than FDS across all configurations. Notably, FDS is also robust to window and hop length. To balance fidelity and efficiency, we use 25 ms windows and a $1/4$ hop, resulting in a hop length of 6.25 ms.

\textbf{FDS Enables GPU Parallelization.} Because solutions for $H_{\text{tract}}$ are independent for each frequency $\omega$, FDS can be efficiently parallelized on a GPU.
We compare the forward runtimes for FDS and TDS on an RTX 4090 with a sampling rate of 16 kHz and $N=32$ sections in Figure ~\ref{fig:runtime}. Given a glottal pulse $u_\text{in}$, TDS takes $1.08$ seconds to synthesize a $1$-second utterance. However, in the frequency domain, this takes only $14.2$ ms.
The runtime tends to increase linearly as the number of windows in the time-domain lattice grows.
On average, FDS is 70x faster with a $25$ ms window and $1/4$ hop.

\section{Experiments}
\label{sec:vocal_recon}
We now describe how our differentiable simulator can be used for solving the speech-to-geometry inverse problem. Because the mapping $\Phi(u_{\text{in}}, A_\theta) \mapsto p_{\text{out}}$ is non-convex, classical reconstruction methods require techniques like second-order optimizers, regularizers, and discrete codebooks to find good solutions~\citep{Panchapagesan2011Chain}. However, inspired by recent work in neural rendering~\citep{mildenhall2020nerf,sitzmann2020implicit}, we discover that we can simply use first-order gradient descent without regularizers by parameterizing $A_\theta$ with a \textit{neural network}. Learning from this, we propose three neural network parameterizations for the vocal tract:

\begin{enumerate}[leftmargin=10pt]
    \item \textbf{\textit{A neural field}}, which demonstrates how parameterizing the area function with a neural network can escape the local minima which trap discrete optimization.
    \item \textbf{\textit{A self-supervised autoencoder}}, that encodes an unlabeled speech sample into an area function, then decodes it back to speech with our simulator. 
    \item \textbf{\textit{A GAN}}, used to synthesize realistic MRI videos of a person while speaking. To our knowledge, this is the first method to infer a MRI video from speech audio without paired speech-MRI data. 
\end{enumerate}
 Each enables novel applications in language and medical imaging. For results of over 30 reconstructed utterances, area functions, MRIs, and singing samples, please find our videos on our \href{https://people.csail.mit.edu/echen/vocal_recon/}{website}.
\begin{figure}[t]
\centering
\includegraphics[width=0.8\linewidth]{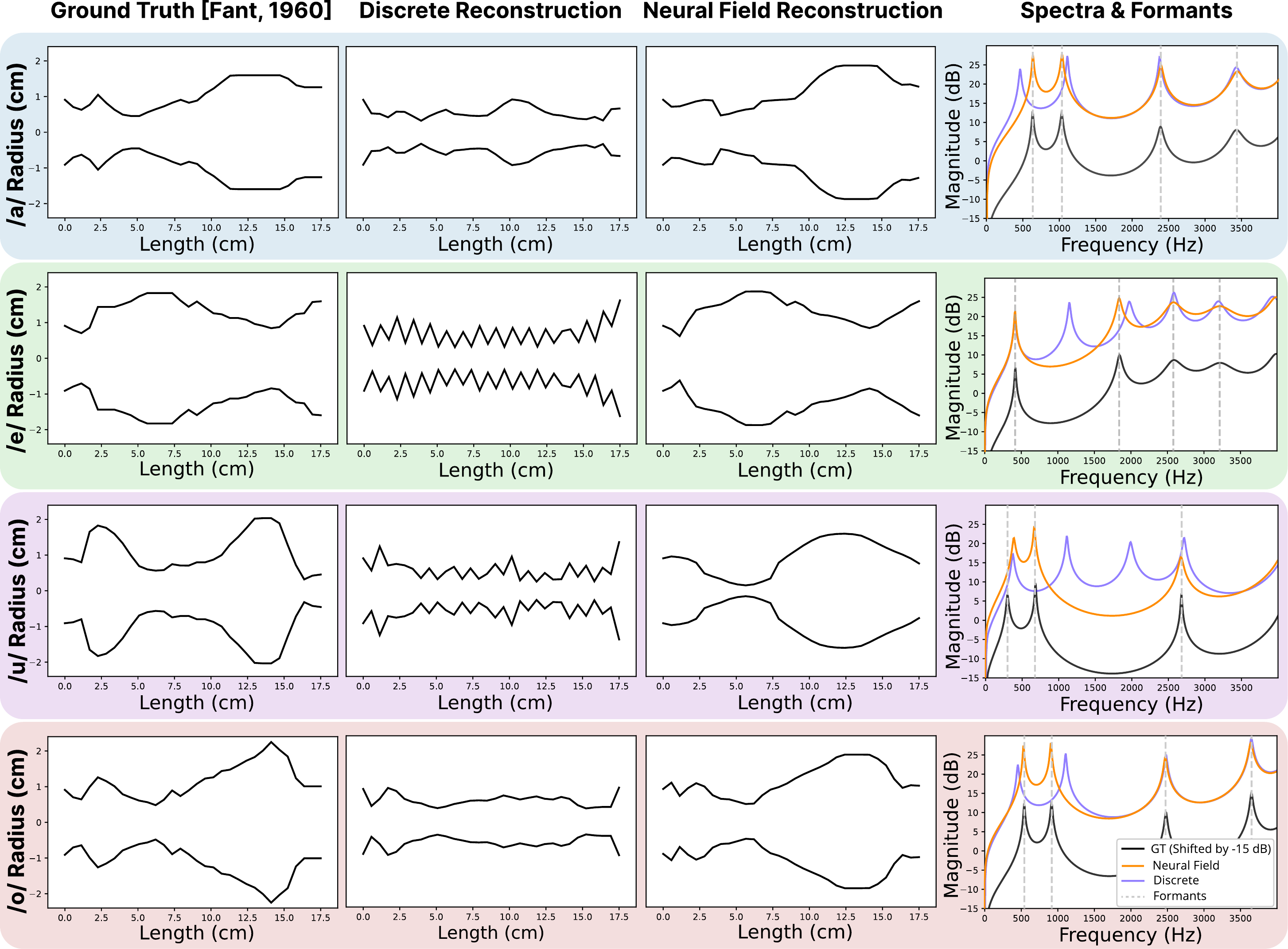}
\caption{\textbf{Neural Fields Smoothly Reconstruct Acoustic Tubes.} Directly optimizing a discrete acoustic tube fails to fit basic vowels. Optimization is trapped around the initialization, and the recovered geometries sometimes suffer from high-frequency oscillation. In comparison, neural fields couple the geometry of each section with shared weights leading to smooth reconstructions.}
\label{fig:neural_field}
\end{figure}

\subsection{Neural Networks Escape Sub-optimality }
\label{subsec:neural_fields}

\textbf{Neural Fields Break Gradient Deadlock.} As common in classical iterative optimization, we initially attempted to directly optimize a discrete array representing the area function $A_\theta[n, k]$ across $N$ spatial segments and $K$ time windows. However, this approach proved unsuccessful. Upon further investigation, we find that parameterizing $A_\theta(x, t)$ with a neural network regularizes the optimization space and better fits high-frequency detail, thereby enabling efficient optimization.
Consider, for instance, the example shown in Figure~\ref{fig:neural_field}. In this example, we initialize the model with a uniform tube and employ the Adam optimizer~\citep{Kingma2014AdamAM} to fit $A_\theta$ to four distinct vowel shapes.
The ground truth is from the X-ray data of~\citet{Fant1960}. When using discrete reconstruction, the optimization stalls as the solutions are often stuck in oscillations around the initial state. It is almost as if the individual $A_n$ sections of the tube are competing against one another with conflicting gradient directions. Conversely, with a continuous neural field, $A_\theta = F_{\theta}(x, t)$, the model is able to recover the general morphology of the area function. Even when the recovered shapes are not flawless, the formants align accurately with the ground truth. We attribute this outcome to the fact that neural fields, and neural networks more generally, share weights between the spatial sections, allowing optimization to succeed in a coarse-to-fine manner. We test two neural fields for $F_\theta$: a fully-connected network with random Fourier feature (RFF) encoding~\citep{tancik2020fourfeat}, and a multiplicative filter network (MFN)~\citep{Fathony2021MultiplicativeFN}. The results in Table~\ref{tab:fant_mse} show that both neural fields significantly outperform the discrete baseline. 

 \begin{table}[h]
\centering
\vspace{-5pt}

\caption[Cross-sectional area reconstruction]{\textbf{Area Reconstruction.} We report the mean squared error of each area parameterization in $\text{cm}^4$. Both neural field methods, RFF and MFN, significantly outperform the discrete baseline.}
\scalebox{0.85}{
\begin{tabular}{lccccccc}
\toprule 
& /\textipa{A}/ & /e/ & /i/ & /\textipa{1}/ & /o/ & /u/  \\ 
\midrule
Discrete 
& 14.66 {\scriptsize $\pm$2.25} 
& 26.22 {\scriptsize $\pm$1.36} 
& 27.66 {\scriptsize $\pm$2.04} 
& 29.65 {\scriptsize $\pm$2.34} 
& 27.42 {\scriptsize $\pm$3.60} 
& 28.19 {\scriptsize $\pm$2.54} \\

RFF 
& \textbf{1.96} {\scriptsize $\pm$4.17} 
& \textbf{1.29} {\scriptsize $\pm$0.90} 
& \textbf{2.46} {\scriptsize $\pm$4.01} 
& \textbf{0.47} {\scriptsize $\pm$0.32} 
& \textbf{1.63} {\scriptsize $\pm$0.40} 
& 27.26 {\scriptsize $\pm$22.30} \\

MFN 
& 5.06 {\scriptsize $\pm$3.64} 
& 15.00 {\scriptsize $\pm$7.06} 
& 5.96 {\scriptsize $\pm$1.37} 
& 5.79 {\scriptsize $\pm$10.52} 
& 8.83 {\scriptsize $\pm$3.22} 
& \textbf{19.98} {\scriptsize $\pm$5.04} \\
\bottomrule
\end{tabular}
}
\label{tab:fant_mse}
\end{table}

\subsection{Self-supervised Autoencoding of Speech}
\begin{table}[t]
\centering
\small
\caption{\textbf{Speech Autoencoding Metrics.} We report WER, CER and ID metrics, with their 95$\%$ confidence intervals, across 11 languages. Our self-supervised autoencoder better encodes intelligible utterances than TensorTract2, an autoencoder trained on synthetic consonant-vowel clusters.}
\scalebox{0.82}{
\begin{tabular}{lcccccccc}
\toprule
 & \multicolumn{2}{c}{\textbf{Ground Truth}} & \multicolumn{3}{c}{\textbf{TensorTract2}~\citep{Krug2025TT2}} & \multicolumn{3}{c}{\textbf{Ours}} \\
 & WER & CER & WER & CER & Speaker ID &  WER & CER & Speaker ID\\ 
\midrule
English  & 3.91 {\scriptsize $\pm$0.84} & 2.05 {\scriptsize $\pm$0.70} & 13.00 {\scriptsize $\pm$1.82} & 7.37 {\scriptsize $\pm$1.11} & 16.86 {\scriptsize $\pm$1.01} & \textbf{5.30} {\scriptsize $\pm$0.94} & \textbf{2.82} {\scriptsize $\pm$0.75} & \textbf{52.0} {\scriptsize $\pm$1.39} \\
German   & 5.53 {\scriptsize $\pm$0.78} & 2.40 {\scriptsize $\pm$0.47} & 17.59 {\scriptsize $\pm$1.15} & 10.85 {\scriptsize $\pm$0.73} &  14.07 {\scriptsize $\pm$1.42}& \textbf{12.49} {\scriptsize $\pm$1.54} & \textbf{7.89} {\scriptsize $\pm$0.74}  & \textbf{54.9} {\scriptsize $\pm$1.45}\\
Dutch    & 9.24 {\scriptsize $\pm$0.87} & 2.59 {\scriptsize $\pm$0.30} & 27.08 {\scriptsize $\pm$1.63} & 12.89 {\scriptsize $\pm$0.91} & 26.97 {\scriptsize $\pm$1.15} & \textbf{17.19} {\scriptsize $\pm$1.27} & \textbf{7.77} {\scriptsize $\pm$0.65} & \textbf{59.3} {\scriptsize $\pm$1.17} \\
French   & 4.72 {\scriptsize $\pm$0.79} & 2.27 {\scriptsize $\pm$0.59} & 19.65 {\scriptsize $\pm$1.71} & 11.37 {\scriptsize $\pm$1.17} & 15.11 {\scriptsize $\pm$1.34} & \textbf{12.91} {\scriptsize $\pm$1.38} & \textbf{7.69} {\scriptsize $\pm$0.85}  & \textbf{47.4} {\scriptsize $\pm$1.46}\\
Spanish  & 3.35 {\scriptsize $\pm$0.69} & 1.37 {\scriptsize $\pm$0.44} & 9.71 {\scriptsize $\pm$0.93}  & 5.13 {\scriptsize $\pm$0.44}  & 14.74 {\scriptsize $\pm$1.49} & \textbf{9.27} {\scriptsize $\pm$1.25} & \textbf{4.43} {\scriptsize $\pm$0.59}  & \textbf{52.9} {\scriptsize $\pm$1.81}\\
Italian  & 9.96 {\scriptsize $\pm$1.18} & 2.28 {\scriptsize $\pm$0.32} & 27.61 {\scriptsize $\pm$1.96} & 10.10 {\scriptsize $\pm$0.95} & 10.34  {\scriptsize $\pm$1.21} & \textbf{22.01} {\scriptsize $\pm$1.94} & \textbf{6.98} {\scriptsize $\pm$0.70}  & \textbf{53.9} {\scriptsize $\pm$1.44}\\
Portuguese & 6.23 {\scriptsize $\pm$0.79} & 2.43 {\scriptsize $\pm$0.44} & 26.61 {\scriptsize $\pm$2.50} & 14.15 {\scriptsize $\pm$2.40} & 17.32 {\scriptsize $\pm$0.99} & \textbf{23.30} {\scriptsize $\pm$9.33} & \textbf{12.68} {\scriptsize $\pm$5.90}  & \textbf{49.1} {\scriptsize $\pm$1.24}\\
Polish   & 4.14 {\scriptsize $\pm$0.63} & 0.98 {\scriptsize $\pm$0.25} & 20.95 {\scriptsize $\pm$1.30} & 7.44 {\scriptsize $\pm$0.50}  & 18.67 {\scriptsize $\pm$0.88} & \textbf{12.88} {\scriptsize $\pm$1.08} & \textbf{4.79} {\scriptsize $\pm$0.42}  & \textbf{49.6} {\scriptsize $\pm$2.43}\\
Korean   & 8.00 {\scriptsize $\pm$2.42} & 2.10 {\scriptsize $\pm$0.79} & 32.32 {\scriptsize $\pm$4.27} & 14.58 {\scriptsize $\pm$2.42} & 23.51 {\scriptsize $\pm$0.77} & \textbf{19.45} {\scriptsize $\pm$3.49} & \textbf{6.05} {\scriptsize $\pm$1.25}  & \textbf{63.4} {\scriptsize $\pm$0.94}\\
Chinese  & -- & 13.54 {\scriptsize $\pm$3.63} & -- & 53.27 {\scriptsize $\pm$42.97} &17.06 {\scriptsize $\pm$1.12}  & -- & \textbf{30.11} {\scriptsize $\pm$13.0}   & \textbf{47.9} {\scriptsize $\pm$1.52}\\
Japanese & -- & 6.89 {\scriptsize $\pm$1.19}  & -- & 15.01 {\scriptsize $\pm$2.38}  &23.76 {\scriptsize $\pm$1.41} & -- & \textbf{11.88} {\scriptsize $\pm$1.57}  & \textbf{53.9} {\scriptsize $\pm$1.44} \\
\bottomrule
\end{tabular}

}
\label{tab:wer}
\end{table}

Since the process of fitting a new neural field to each utterance is computationally expensive, we also train an encoder $\mathcal{E}_\theta: p_{\text{tgt}} \mapsto A$ that maps a speech recording directly to an area function. The area function is then decoded back to speech with our simulator, $\Phi$.  We use the Wav2Vec 2.0 architecture~\citep{Baevski2020wav2vec2A} for $\mathcal{E}_\theta$. The encoder is trained end-to-end with the loss $\mathcal{L}_2(p_{\text{tgt}}, \Phi (u_{\text{in}}, \mathcal{E}_\theta(p_{\text{tgt}}))$. No paired data between $p_{\text{tgt}}$ and $A(x, t)$ is required since $\Phi$ itself physically constrains the latent space of the autoencoder to be plausible. Note that unlike prior articulatory autoencoders~\citep{Krug2025TT2,Cho2024EMA} that require paired speech and articulatory data, our model is self-supervised.

\textbf{Analysis-by-Synthesis Enables Scalable Training.} 
While there is no prior self-supervised articulatory autoencoder to compare to, we can still evaluate our work against TensorTract2 (TT2)~\citep{Krug2025TT2}, an articulatory autoencoder trained on paired data. TT2 is trained on synthetic consonant-vowel clusters generated by VocalTractLab~\citep{BirkholzVTL}. 
In comparison, our autoencoder benefits from training on real-world speech datasets because it is self-supervised. Described in Appendix~\ref{app:autoencoding}, we train our autoencoder on 11 different languages. 200 synthesized utterances from the test set of each language are then input to Whisper~\citep{Radford2022Whisper} to evaluate their word error rates (WER) and character error rates (CER). We also report cosine similarity between ECAPA-TDNN~\citep{desplanques2020ecapa} speaker identity embeddings. Across 11 languages, Table~\ref{tab:wer} demonstrates that our simulator outputs utterances that are more intelligible and better preserve identity.
\label{subsec:autoencoding}
\begin{wrapfigure}{U}{.25\textwidth}
\vspace{-15pt}
    \includegraphics[width=\linewidth]{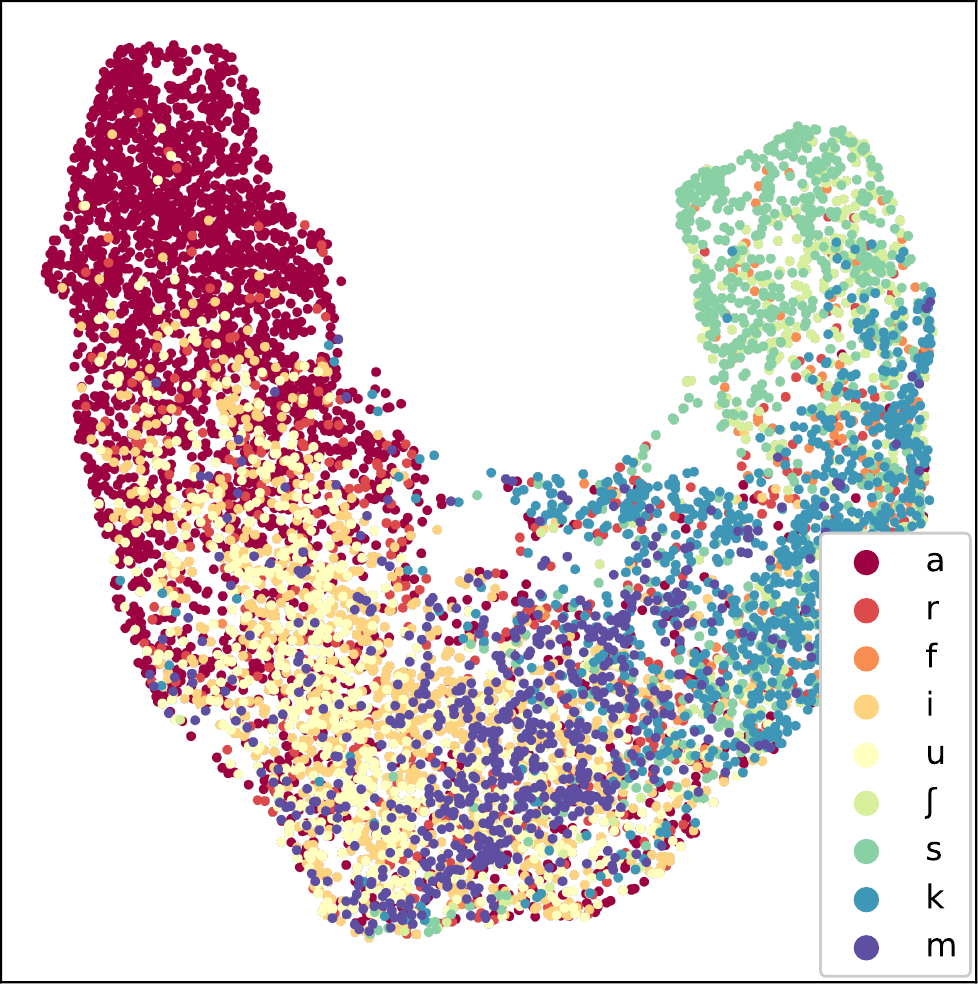}
    \caption{Encoded areas form a spectrum.}
    \label{fig:phoneme_umap}
    \vspace{-10pt}
\end{wrapfigure}

\textbf{Vocal Tract Area Functions Encode a Linguistic Spectrum.} Consisting of only 32 spatial sections,  area functions are an exceptionally compact representation of speech. To investigate what linguistic information they retain---if any---we plot a UMAP~\citep{mcinnes2018umap-software} projection of the area functions and color each point by its phoneme (Figure~\ref{fig:phoneme_umap}). Despite receiving no phonetic labels during training, the area functions remarkably form a continuous spectrum of phonemes, ordered by place of articulation. On the left are vowels like /a/, /i/, and /u/, while constrictions like /s/ and /k/ occupy the right. Although phonemes are often thought of as discrete symbols, our encoder naturally embeds them in a low-dimensional continuous space. This is a fully anticipated phenomenon, considering that the transitions in oral structure between phonemes occur smoothly in spoken language.

\subsection{Physically-Grounded MRI Visualization from Speech}
A visual representation of how one's lip position, tongue geometry, etc. affect their speech would be useful in applications such as language learning and singing instruction. However, directly capturing this data with an MRI machine is impractical outside of specialized settings. Our differentiable simulator thus enables a task that, to our knowledge, has not previously been attempted: visualizing an MRI of a person's vocal tract from their speech, without any paired speech-MRI training data.
\begin{figure*}[t]
    \centering
    \includegraphics[width=\linewidth]{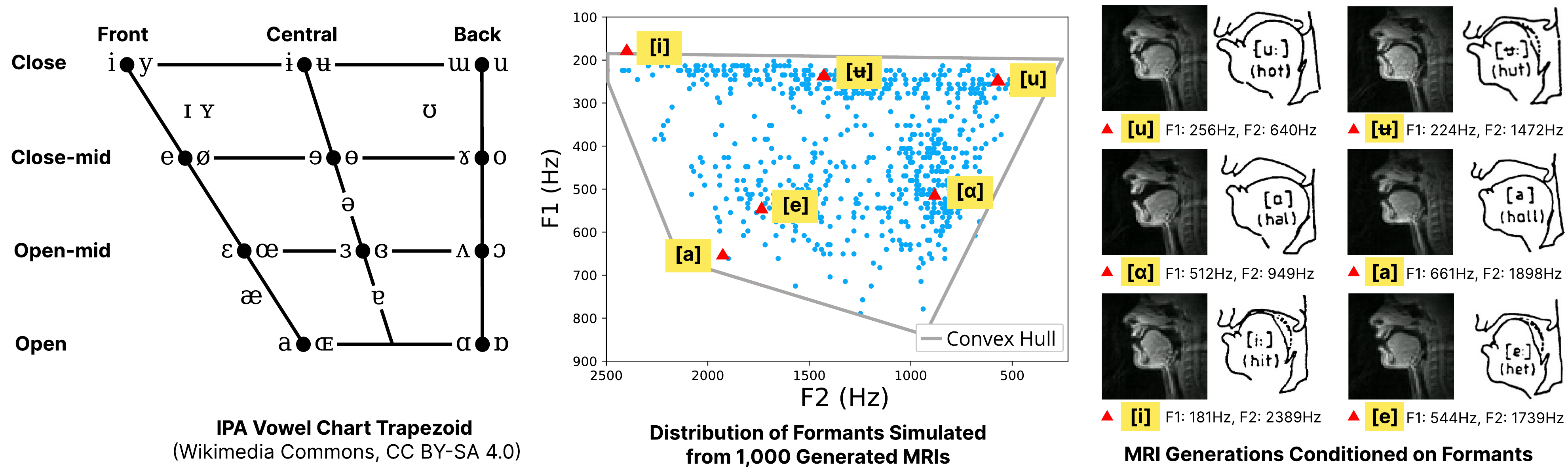}
    \caption{\textbf{Phonetic Structure Emerges without Supervision.} We randomly generate 1,000 MRIs from our GAN, then compute their formants with our simulator. The convex hull of formants resemble a trapezoid corresponding almost exactly to the IPA vowel chart. The anatomy of each generated MRI corresponds closely to textbook illustrations of vowels from~\citet{Fant2004SpeechAA}. }
    \label{fig:formant_space}
\end{figure*}

Our method is shown in Figure~\ref{fig:recon}. First, we pre-train a StyleGAN2 network~\citep{Karras2019AnalyzingAI} on MRI scans from~\citep{Lim2021}. The trained GAN $G$ maps a latent vector $w\in \mathbb{R}^{512}$ to an MRI $I$. For reconstruction, we start by initializing a sequence of latents, $\theta = w_1, \dots, w_K$, corresponding to a video $I_1, \dots, I_K = G(w_1), \dots, G(w_K)$. We then extract the cross-sectional area from each $I$, denoted with $\psi(I_k)=  A[1\texttt{:}N, k] $. And finally, we run simulation on $A$ to generate an utterance $p_{\text{out}}$. Overall, $
    p_{\text{out}} = \Phi(u_{\text{in}}, \psi(G(w_1)), \dots, \psi(G(w_K) ))$. 
To fit an utterance, we backpropagate the entire process end-to-end, deriving gradients for each $w_k$. We differentiably segment the cross-sectional areas from each MRI frame by first softly binarizing it with a sigmoid function, then extracting the areas along a guide. The guide, pictured in orange in Figure~\ref{fig:recon}, simply consists of a quarter circle arc and a line segment, connecting the glottis to the lips. Along each orange line, the cross-sectional areas are approximated by summing the binarized pixels. 
Because this technique does not require paired speech-MRI data it is fully self-supervised.

\textbf{Phonetic Structure Emerges without Supervision.} We explore how physically grounded the audio-visual representation is. In Figure~\ref{fig:formant_space}, we visualize the formant space generated from our learned representation. We randomly sample 1,000 MRI frames from the GAN, use our simulator to synthesize vowels extracted from their cross-sectional areas, then plot their first two formants. Despite the absence of any explicit supervision, the formant space's convex hull forms a trapezoid almost exactly resembling the trapezoidal vowel chart from the International Phonetic Alphabet (IPA) chart. When compared to vowel examples from~\citep{Fant2004SpeechAA}, we observe a very close correspondence in tongue positioning as well. The fact that the IPA emerges spontaneously from this audio-visual representation gives us confidence that our simulator captures geometrically meaningful vocal tract configurations.

\textbf{Acoustic Grounding Prevents Video Collapse.}\begin{wrapfigure}{U}{.26\textwidth}
    \centering
    \vspace{-8pt}
    \captionof{table}{\textbf{MRI Metrics}}
    \label{tab:mri_metrics}
    \scalebox{0.78}{
        \begin{tabular}{l|cc}
            \toprule
            & \textbf{Ours} & $\substack{\text{\large Speech} \\ \text{\large 2rtMRI}}$ \\
            \midrule
            \textbf{Audio} & & \\
            SI-SDR $\uparrow$   & \textbf{13.08}  & N/A \\
            STOI  $\uparrow$ & \textbf{0.92} & N/A \\
            PESQ $\uparrow$  & \textbf{2.09}  & N/A \\
            \midrule
            \textbf{Visual} & & \\
            FVD $\downarrow$ & \textbf{623}   & 2949 \\ 
            SSIM $\uparrow$ & \textbf{0.352} & 0.317 \\
            LPIPS $\downarrow$ & \textbf{0.159} & 0.355 \\
            \bottomrule
        \end{tabular}
    }
    \vspace{-5pt}
\end{wrapfigure}
 Without a physical model to connect geometry to speech, prior works that tackle this task act as black boxes. For example, Speech2rtMRI~\citep{Nguyen2024Speech2rtMRISD}, the closest prior work, trains a diffusion model on paired data between speech and MRI. The quality of this diffusion model is limited by MRI data, which is difficult to collect. Furthermore, because Speech2rtMRI does not leverage a physical representation, the generated videos may lack physical plausibility. It can only generate 10-20 frames of video. Our ``analysis by synthesis'' technique on the other hand makes MRI visualization tractable for the first time by physically constraining the video generations with audio. We compare our methods on 40 sequences of 3 seconds, corresponding to 9,900 video frames. 

\begin{figure}[t] 
    \centering
    
        \centering
        \includegraphics[width=0.49\linewidth]{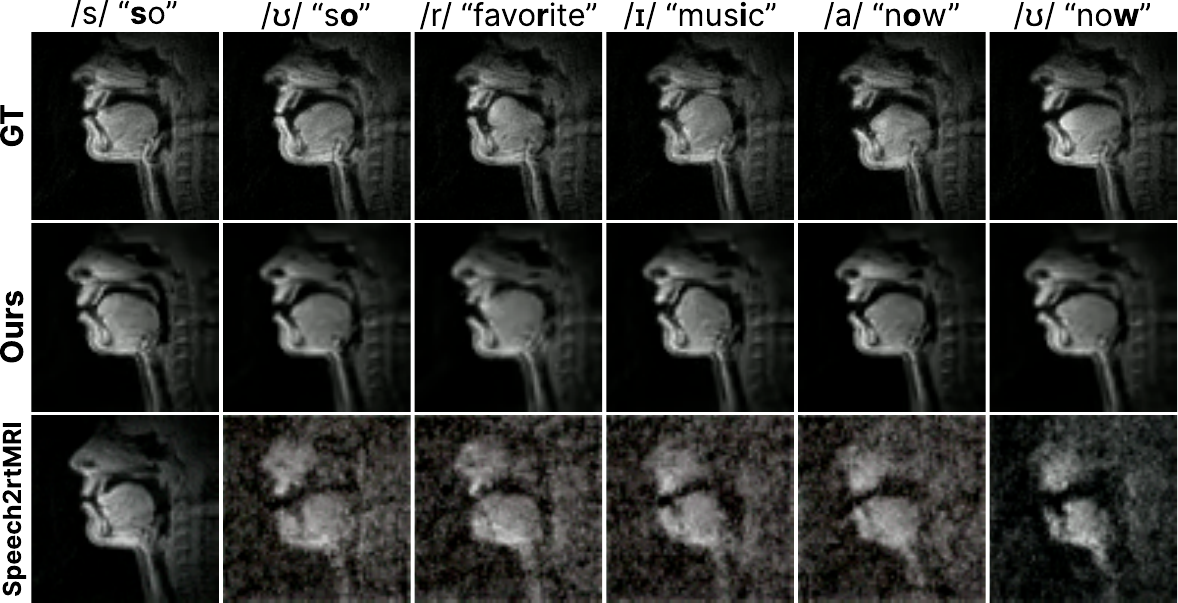}
        \includegraphics[width=0.49\linewidth]{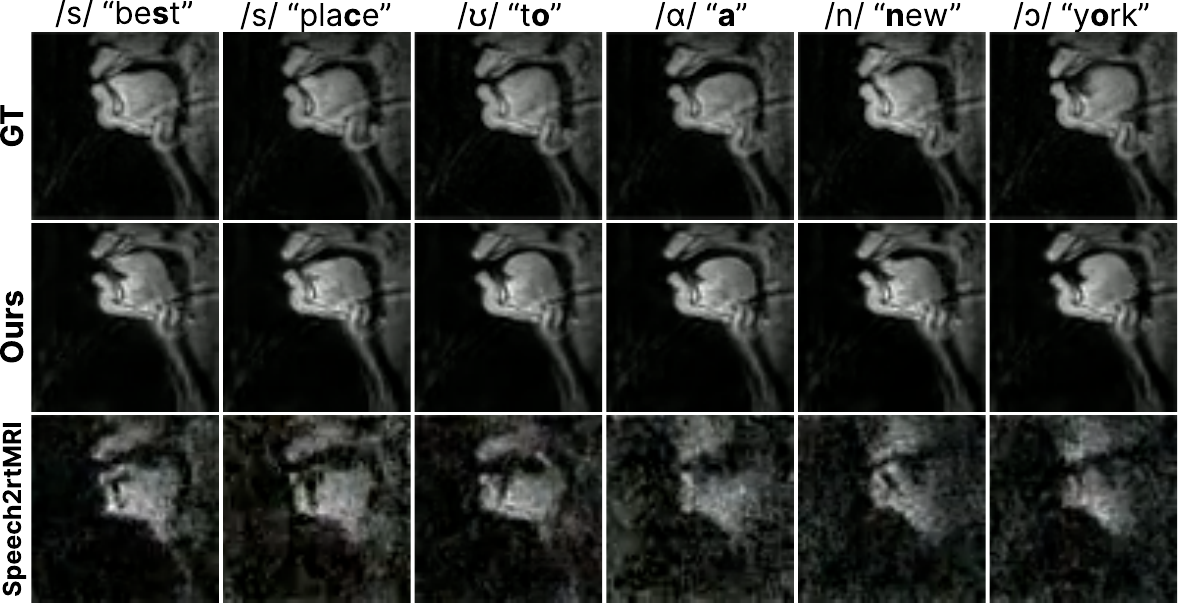}

        \caption{\textbf{Acoustic Grounding Prevents Video Collapse.} Because Speech2rtMRI~\citep{Nguyen2024Speech2rtMRISD} does not leverage a physical representation, it tends to degenerate over time. Our method remains stable. }
        \label{fig:mri_recon}

\end{figure}

We evaluate each method on their acoustic and visual quality. According to the experimental results, Speech2rtMRI suffers from noise degradation. As illustrated in Figure~\ref{fig:mri_recon}, the diffusion model collapses and the generated images exhibit severe noise and unrealistic artifacts. The quantitative results show that the diffusion model achieves a Fr\'{e}chet Video Distance (FVD) of 2949, an order of magnitude higher than our value of 623. Despite begin trained on unlabeled data, our method still outperforms Speech2rtMRI on SSIM (structural similarity index measure)  and LPIPS (learned perceptual image patch similarity)~\citep{Zhang2018TheUE} as well. These two image metrics measure how accurately the MRI's structure matches the ground truth. Our method also simultaneously simulates audio. Despite the challenging task of backpropagating through an MRI generator, it is still able to produce intelligible audio, something not previously possible.

\section{Conclusion}
We have introduced a differentiable and GPU parallelizable physics-based model of human speech. We have discovered several technical insights along the way: (1) Modeling acoustics in the frequency domain enables stable gradient flow, and is 70x more GPU parallelizable than finite differences in time. (2) We introduce a differentiable model for consonant formation. (3) Parameterizing the area functions with a neural field alleviates the sub-optimality failures common in acoustic inverse problems. Beyond the simulator itself, we enable two applications that were not previously possible. (1) We train a self-supervised autoencoder that predicts the vocal tract's area function directly from raw audio, without any paired articulatory data. (2) We provide a way to visualize an MRI video of one's vocal tract from their speech, without paired speech-MRI training data.

\textbf{Limitations.} We note several directions for future improvement. As shown in Figure~\ref{fig:mri_recon}, our MRI images sometimes do not correspond directly to the ground truth. Part of the reason is that the inverse problem is inherently ill-posed: different geometries can produce the same sound. Furthermore, the MRI dataset used does not image the nasal tract. Reconstructing the nasal tract would produce more faithful formants and anti-formants.

\textbf{Broader Impacts.} While this work is primarily focused on acoustics, we believe that this work opens the door to exciting applications beyond acoustics. Our acoustic simulator can be used, for instance, in music, as a tool for voice coaching or singing synthesis; in cognitive science, such as the study of language acquisition~\citep{beguvs2023articulation}; and in medicine, such as speech pathology. Another direction for future work is to extend our model to other animals, such as sperm whales~\citep{Sharma2023ContextualAC} or zebra finches.

\newpage
\section*{Acknowledgments}
This work would not have been possible without discussions with Morris Alper about phonetics; with Mark Rau about acoustic simulation; and with Matthew Caren and Kartik Chandra about their work on vocal imitation. 
EC is supported by the NSF Graduate Research Fellowship Program. 

Special thanks to our foreign language contributors: Cl\'{e}ment Jambon for French, David Charatan for German,  Chonghyuk (Andrew) Song for Korean, and Emerald Liu for Chinese. 

\bibliography{bibliography}

\begin{thebibliography}{62}
\providecommand{\natexlab}[1]{#1}
\providecommand{\url}[1]{\texttt{#1}}
\expandafter\ifx\csname urlstyle\endcsname\relax
  \providecommand{\doi}[1]{doi: #1}\else
  \providecommand{\doi}{doi: \begingroup \urlstyle{rm}\Url}\fi

\bibitem[Baevski et~al.(2020)Baevski, Zhou, rahman Mohamed, and
  Auli]{Baevski2020wav2vec2A}
Alexei Baevski, Henry Zhou, Abdel rahman Mohamed, and Michael Auli.
\newblock wav2vec 2.0: A framework for self-supervised learning of speech
  representations.
\newblock \emph{ArXiv}, abs/2006.11477, 2020.
\newblock URL \url{https://api.semanticscholar.org/CorpusID:219966759}.

\bibitem[Begu{\v{s}} et~al.(2023)Begu{\v{s}}, Zhou, Wu, and
  Anumanchipalli]{beguvs2023articulation}
Ga{\v{s}}per Begu{\v{s}}, Alan Zhou, Peter Wu, and Gopala~K Anumanchipalli.
\newblock Articulation gan: Unsupervised modeling of articulatory learning.
\newblock In \emph{ICASSP 2023-2023 IEEE International Conference on Acoustics,
  Speech and Signal Processing (ICASSP)}, pages 1--5. IEEE, 2023.

\bibitem[Bilbao(2009)]{BilbaoNumerical2009}
Stefan Bilbao.
\newblock \emph{Acoustic Tubes}, chapter~9, pages 249--286.
\newblock John Wiley \& Sons, Ltd, 2009.
\newblock ISBN 9780470749012.
\newblock \doi{https://doi.org/10.1002/9780470749012.ch9}.

\bibitem[Birkholz(2013)]{BirkholzVTL}
Peter Birkholz.
\newblock Modeling consonant-vowel coarticulation for articulatory speech
  synthesis.
\newblock \emph{PLOS ONE}, 8\penalty0 (4):\penalty0 1--17, 04 2013.
\newblock \doi{10.1371/journal.pone.0060603}.
\newblock URL \url{https://doi.org/10.1371/journal.pone.0060603}.

\bibitem[Birkholz and Jackel(2004)]{Birkholz2004InfluenceOT}
Peter Birkholz and Dietmar Jackel.
\newblock Influence of temporal discretization schemes on formant frequencies
  and bandwidths in time domain simulations of the vocal tract system.
\newblock In \emph{Interspeech}, 2004.
\newblock URL \url{https://api.semanticscholar.org/CorpusID:15404079}.

\bibitem[Birkholz and Jackel(2006)]{Birkholz2006NoiseSA}
Peter Birkholz and Dietmar Jackel.
\newblock Noise sources and area functions for the synthesis of fricative
  consonants.
\newblock 2006.
\newblock URL \url{https://api.semanticscholar.org/CorpusID:12604988}.

\bibitem[Bridson et~al.(2007)Bridson, Houriham, and
  Nordenstam]{Bridson2007CurlnoiseFP}
Robert Bridson, Jim Houriham, and Marcus Nordenstam.
\newblock Curl-noise for procedural fluid flow.
\newblock \emph{ACM SIGGRAPH 2007 papers}, 2007.
\newblock URL \url{https://api.semanticscholar.org/CorpusID:10174968}.

\bibitem[Caren et~al.(2024)Caren, Chandra, Tenenbaum, Ragan-Kelley, and
  Ma]{Caren2024}
Matthew Caren, Kartik Chandra, Joshua Tenenbaum, Jonathan Ragan-Kelley, and
  Karima Ma.
\newblock Sketching with your voice: "non-phonorealistic" rendering of sounds
  via vocal imitation.
\newblock In \emph{SIGGRAPH Asia 2024 Conference Papers}, SA '24, New York, NY,
  USA, 2024. Association for Computing Machinery.
\newblock URL \url{https://doi.org/10.1145/3680528.3687679}.

\bibitem[Cho et~al.(2024)Cho, Wu, Prabhune, Agarwal, and
  Anumanchipalli]{Cho2024EMA}
Cheol~Jun Cho, Peter Wu, Tejas~S. Prabhune, Dhruv Agarwal, and Gopala~K.
  Anumanchipalli.
\newblock Coding speech through vocal tract kinematics.
\newblock \emph{IEEE Journal of Selected Topics in Signal Processing},
  18\penalty0 (8):\penalty0 1427--1440, 2024.
\newblock \doi{10.1109/JSTSP.2024.3497655}.

\bibitem[Chodroff et~al.(2024)Chodroff, Pazon, Baker, and
  Moran]{Chodroff2024voxangeles}
Eleanor Chodroff, Blaz Pazon, Annie Baker, and Steven Moran.
\newblock Phonetic segmentation of the ucla phonetics lab archive.
\newblock In \emph{International Conference on Language Resources and
  Evaluation}, 2024.
\newblock URL \url{https://api.semanticscholar.org/CorpusID:268732610}.

\bibitem[Chung et~al.(2023)Chung, Kim, Mccann, Klasky, and
  Ye]{chung2023diffusion}
Hyungjin Chung, Jeongsol Kim, Michael~Thompson Mccann, Marc~Louis Klasky, and
  Jong~Chul Ye.
\newblock Diffusion posterior sampling for general noisy inverse problems.
\newblock In \emph{The Eleventh International Conference on Learning
  Representations}, 2023.
\newblock URL \url{https://openreview.net/forum?id=OnD9zGAGT0k}.

\bibitem[Deller~Jr et~al.(1993)Deller~Jr, Proakis, and
  Hansen]{deller1993discrete}
John~R Deller~Jr, John~G Proakis, and John~H Hansen.
\newblock \emph{Discrete time processing of speech signals}.
\newblock Prentice Hall PTR, 1993.

\bibitem[Desplanques et~al.(2020)Desplanques, Thienpondt, and
  Demuynck]{desplanques2020ecapa}
Brecht Desplanques, Jenthe Thienpondt, and Kris Demuynck.
\newblock {ECAPA-TDNN: Emphasized Channel Attention, propagation and
  aggregation in TDNN based speaker verification}.
\newblock In \emph{Interspeech 2020}, pages 3830--3834, 2020.

\bibitem[Dunn(1950)]{dunn1950calculation}
Hugh~K Dunn.
\newblock The calculation of vowel resonances, and an electrical vocal tract.
\newblock \emph{The Journal of the Acoustical Society of America}, 22\penalty0
  (6):\penalty0 740--753, 1950.

\bibitem[Engel et~al.(2020)Engel, Hantrakul, Gu, and Roberts]{engelddsp}
Jesse Engel, Lamtharn Hantrakul, Chenjie Gu, and Adam Roberts.
\newblock Ddsp: Differentiable digital signal processing.
\newblock \emph{ArXiv}, abs/2001.04643, 2020.
\newblock URL \url{https://api.semanticscholar.org/CorpusID:210473083}.

\bibitem[Fant(1960)]{Fant1960}
Gunnar Fant.
\newblock \emph{Acoustic theory of speech production, with calculations based
  on X-ray studies of Russian articulations.}
\newblock Mouton and Co. N.V., The Hague, 1960.

\bibitem[Fant(2004)]{Fant2004SpeechAA}
Gunnar Fant.
\newblock \emph{Speech Acoustics and Phonetics}.
\newblock Text, Speech and Language Technology. Springer Dordrecht, 2004.
\newblock \doi{10.1007/978-1-4020-5746-5}.
\newblock URL \url{https://api.semanticscholar.org/CorpusID:60121294}.

\bibitem[Fant et~al.(1985)Fant, Liljencrants, and Lin]{fant1985four}
Gunnar Fant, Johan Liljencrants, and Qi-guang Lin.
\newblock A four-parameter model of glottal flow.
\newblock \emph{STL-QPSR}, 4\penalty0 (1985):\penalty0 1--13, 1985.

\bibitem[Fathony et~al.(2021)Fathony, Sahu, Willmott, and
  Kolter]{Fathony2021MultiplicativeFN}
Rizal Fathony, Anit~Kumar Sahu, Devin Willmott, and J.~Zico Kolter.
\newblock Multiplicative filter networks.
\newblock In \emph{International Conference on Learning Representations}, 2021.
\newblock URL \url{https://api.semanticscholar.org/CorpusID:235613628}.

\bibitem[Finnendahl et~al.(2025)Finnendahl, Worchel, J{\"u}terbock, Wujecki,
  Brinkmann, Weinzierl, and Alexa]{finnendahl2025differentiable}
Ugo Finnendahl, Markus Worchel, Tobias J{\"u}terbock, Daniel Wujecki, Fabian
  Brinkmann, Stefan Weinzierl, and Marc Alexa.
\newblock Differentiable geometric acoustic path tracing using time-resolved
  path replay backpropagation.
\newblock \emph{ACM Transactions on Graphics (TOG)}, 44\penalty0 (4), 2025.
\newblock \doi{10.1145/3730900}.

\bibitem[Flanagan(1972)]{Flanagan1972}
James~L. Flanagan.
\newblock \emph{Speech Analysis Synthesis and Perception}.
\newblock Communication and Cybernetics. Springer Berlin, Heidelberg, 2
  edition, 1972.
\newblock \doi{10.1007/978-3-662-01562-9}.

\bibitem[Freeman et~al.(2021)Freeman, Frey, Raichuk, Girgin, Mordatch, and
  Bachem]{brax2021github}
C.~Daniel Freeman, Erik Frey, Anton Raichuk, Sertan Girgin, Igor Mordatch, and
  Olivier Bachem.
\newblock Brax - a differentiable physics engine for large scale rigid body
  simulation, 2021.
\newblock URL \url{http://github.com/google/brax}.

\bibitem[Glassner(1989)]{glassner1989introduction}
Andrew~S Glassner.
\newblock \emph{An introduction to ray tracing}.
\newblock Morgan Kaufmann, 1989.

\bibitem[Hahn et~al.(1995)Hahn, Geigel, Lee, Gritz, Takala, and
  Mishra]{hahn1995integrated}
James~K Hahn, Joe Geigel, Jong~Won Lee, Larry Gritz, Tapio Takala, and Suneil
  Mishra.
\newblock An integrated approach to motion and sound.
\newblock \emph{The Journal of Visualization and Computer Animation},
  6\penalty0 (2):\penalty0 109--123, 1995.

\bibitem[Hu et~al.(2019)Hu, Anderson, Li, Sun, Carr, Ragan-Kelley, and
  Durand]{Hu2019DiffTaichiDP}
Yuanming Hu, Luke Anderson, Tzu-Mao Li, Qi~Sun, Nathan~A. Carr, Jonathan
  Ragan-Kelley, and Fr{\'e}do Durand.
\newblock Difftaichi: Differentiable programming for physical simulation.
\newblock \emph{ArXiv}, abs/1910.00935, 2019.
\newblock URL \url{https://api.semanticscholar.org/CorpusID:203626832}.

\bibitem[Karras et~al.(2019)Karras, Laine, Aittala, Hellsten, Lehtinen, and
  Aila]{Karras2019AnalyzingAI}
Tero Karras, Samuli Laine, Miika Aittala, Janne Hellsten, Jaakko Lehtinen, and
  Timo Aila.
\newblock Analyzing and improving the image quality of stylegan.
\newblock \emph{2020 IEEE/CVF Conference on Computer Vision and Pattern
  Recognition (CVPR)}, pages 8107--8116, 2019.
\newblock URL \url{https://api.semanticscholar.org/CorpusID:209202273}.

\bibitem[Kelly and Lochbaum(1962)]{kelly_lochbaum_1962}
J.~L.~Jr. Kelly and C.~C. Lochbaum.
\newblock Speech synthesis.
\newblock In \emph{Proceedings of the Stockholm Speech Communication Seminar},
  1962.

\bibitem[Kingma and Ba(2014)]{Kingma2014AdamAM}
Diederik~P. Kingma and Jimmy Ba.
\newblock Adam: A method for stochastic optimization.
\newblock \emph{CoRR}, abs/1412.6980, 2014.
\newblock URL \url{https://api.semanticscholar.org/CorpusID:6628106}.

\bibitem[Koizumi et~al.(2023)Koizumi, Zen, Karita, Ding, Yatabe, Morioka,
  Bacchiani, Zhang, Han, and Bapna]{Koizumi2023LibriTTSRAR}
Yuma Koizumi, Heiga Zen, Shigeki Karita, Yifan Ding, Kohei Yatabe, Nobuyuki
  Morioka, Michiel Bacchiani, Yu~Zhang, Wei Han, and Ankur Bapna.
\newblock Libritts-r: A restored multi-speaker text-to-speech corpus.
\newblock \emph{ArXiv}, abs/2305.18802, 2023.
\newblock URL \url{https://api.semanticscholar.org/CorpusID:258967444}.

\bibitem[Kong et~al.(2020)Kong, Kim, and Bae]{kong2020hifigan}
Jungil Kong, Jaehyeon Kim, and Jaekyoung Bae.
\newblock Hifi-gan: Generative adversarial networks for efficient and high
  fidelity speech synthesis.
\newblock In \emph{NeurIPS}, volume~33, pages 17022--17033, 2020.

\bibitem[Krug et~al.(2025)Krug, Wagner, Birkholz, and Stich]{Krug2025TT2}
Paul~Konstantin Krug, Christoph Wagner, Peter Birkholz, and Timo Stich.
\newblock Precisely controllable neural speech synthesis.
\newblock In \emph{ICASSP 2025 - 2025 IEEE International Conference on
  Acoustics, Speech and Signal Processing (ICASSP)}, pages 1--5, 2025.
\newblock \doi{10.1109/ICASSP49660.2025.10890772}.

\bibitem[Kumar et~al.(2023)Kumar, Tan, Ni, Manocha, Zhang, Henderson, and
  Xu]{Kumar2023TorchaudioSquimRS}
Anurag Kumar, Ke~Tan, Zhaoheng Ni, Pranay Manocha, Xiaohui Zhang, Ethan
  Henderson, and Buye Xu.
\newblock Torchaudio-squim: Reference-less speech quality and intelligibility
  measures in torchaudio.
\newblock \emph{ICASSP 2023 - 2023 IEEE International Conference on Acoustics,
  Speech and Signal Processing (ICASSP)}, pages 1--5, 2023.
\newblock URL \url{https://api.semanticscholar.org/CorpusID:257921409}.

\bibitem[Kuttruff(2016)]{kuttruff2016room}
Heinrich Kuttruff.
\newblock \emph{Room acoustics}.
\newblock Crc Press, 2016.

\bibitem[Lan et~al.(2024)Lan, Zheng, Zheng, and Zhao]{lan2024acoustic}
Zitong Lan, Chenhao Zheng, Zhiwei Zheng, and Mingmin Zhao.
\newblock Acoustic volume rendering for neural impulse response fields.
\newblock In \emph{Advances in Neural Information Processing Systems
  (NeurIPS)}, 2024.

\bibitem[Liang et~al.(2023)Liang, Huang, Tian, Kumar, and Xu]{liang2023av}
Susan Liang, Chao Huang, Yapeng Tian, Anurag Kumar, and Chenliang Xu.
\newblock Av-nerf: Learning neural fields for real-world audio-visual scene
  synthesis.
\newblock \emph{Advances in Neural Information Processing Systems},
  36:\penalty0 37472--37490, 2023.

\bibitem[Lim et~al.(2021)Lim, Toutios, Bliesener, Tian, Lingala, Vaz, Sorensen,
  Oh, Harper, Chen, Lee, Töger, Montesserin, Smith, Godinez, Goldstein, Byrd,
  Nayak, and Narayanan]{Lim2021}
Yongwan Lim, Asterios Toutios, Yannick Bliesener, Ye~Tian, Sajan~Goud Lingala,
  Colin Vaz, Tanner Sorensen, Miran Oh, Sarah Harper, Weiyi Chen, Yoonjeong
  Lee, Johannes Töger, Mairym~Lloréns Montesserin, Caitlin Smith, Bianca
  Godinez, Louis Goldstein, Dani Byrd, Krishna~S Nayak, and Shrikanth
  Narayanan.
\newblock {A multispeaker dataset of raw and reconstructed speech production
  real-time MRI video and 3D volumetric images}.
\newblock 2 2021.
\newblock \doi{10.6084/m9.figshare.13725546.v1}.
\newblock URL
  \url{https://figshare.com/articles/dataset/A_multispeaker_dataset_of_raw_and_reconstructed_speech_production_real-time_MRI_video_and_3D_volumetric_images/13725546}.

\bibitem[Luo et~al.(2022)Luo, Du, Tarr, Tenenbaum, Torralba, and
  Gan]{luo2022learning}
Andrew Luo, Yilun Du, Michael Tarr, Josh Tenenbaum, Antonio Torralba, and
  Chuang Gan.
\newblock Learning neural acoustic fields.
\newblock \emph{Advances in Neural Information Processing Systems},
  35:\penalty0 3165--3177, 2022.

\bibitem[Macklin(2022)]{warp2022}
Miles Macklin.
\newblock Warp: A high-performance python framework for gpu simulation and
  graphics.
\newblock \url{https://github.com/nvidia/warp}, March 2022.
\newblock NVIDIA GPU Technology Conference (GTC).

\bibitem[Maeda(1982)]{Maeda1982}
Shinji Maeda.
\newblock A digital simulation method of the vocal-tract system.
\newblock \emph{Speech Communication}, 1\penalty0 (3):\penalty0 199--229, 1982.
\newblock ISSN 0167-6393.
\newblock \doi{https://doi.org/10.1016/0167-6393(82)90017-6}.
\newblock URL
  \url{https://www.sciencedirect.com/science/article/pii/0167639382900176}.

\bibitem[McInnes et~al.(2018)McInnes, Healy, Saul, and
  Grossberger]{mcinnes2018umap-software}
Leland McInnes, John Healy, Nathaniel Saul, and Lukas Grossberger.
\newblock Umap: Uniform manifold approximation and projection.
\newblock \emph{The Journal of Open Source Software}, 3\penalty0 (29):\penalty0
  861, 2018.

\bibitem[Metzger et~al.(2023)Metzger, Littlejohn, Silva, Moses, Seaton, Wang,
  Dougherty, Liu, Wu, Berger, Zhuravleva, Tu-Chan, Ganguly, Anumanchipalli, and
  Chang]{Metzger2023AHN}
Sean~L. Metzger, Kaylo~T Littlejohn, Alexander~B. Silva, David~Aaron Moses,
  Margaret~P. Seaton, Ran Wang, Maximilian~E. Dougherty, Jessie~R. Liu, Peter
  Wu, Michael Berger, Inga Zhuravleva, Adelyn~P. Tu-Chan, Karunesh Ganguly,
  Gopala~Krishna Anumanchipalli, and Edward~F. Chang.
\newblock A high-performance neuroprosthesis for speech decoding and avatar
  control.
\newblock \emph{Nature}, 620:\penalty0 1037--1046, 2023.
\newblock URL \url{https://api.semanticscholar.org/CorpusID:261098775}.

\bibitem[Mildenhall et~al.(2020)Mildenhall, Srinivasan, Tancik, Barron,
  Ramamoorthi, and Ng]{mildenhall2020nerf}
Ben Mildenhall, Pratul~P Srinivasan, Matthew Tancik, Jonathan~T Barron, Ravi
  Ramamoorthi, and Ren Ng.
\newblock Nerf: Representing scenes as neural radiance fields for view
  synthesis.
\newblock In \emph{European Conference on Computer Vision}, pages 405--421.
  Springer, 2020.

\bibitem[Nguyen et~al.(2024)Nguyen, Foley, Huang, Shi, Feng, and
  Narayanan]{Nguyen2024Speech2rtMRISD}
Hong Nguyen, Sean Foley, Kevin Huang, Xuan Shi, Tiantian Feng, and Shrikanth~S.
  Narayanan.
\newblock Speech2rtmri: Speech-guided diffusion model for real-time mri video
  of the vocal tract during speech.
\newblock \emph{ArXiv}, abs/2409.15525, 2024.
\newblock URL \url{https://api.semanticscholar.org/CorpusID:272832441}.

\bibitem[Nieradzik(2025)]{nieradzik2025swiftf0}
Lars Nieradzik.
\newblock Swiftf0: Fast and accurate monophonic pitch detection, 2025.
\newblock URL \url{https://arxiv.org/abs/2508.18440}.

\bibitem[Oord et~al.(2016)Oord, Dieleman, Zen, Simonyan, Vinyals, Graves,
  Kalchbrenner, Senior, and Kavukcuoglu]{oord2016wavenet}
Aaron van~den Oord, Sander Dieleman, Heiga Zen, Karen Simonyan, Oriol Vinyals,
  Alex Graves, Nal Kalchbrenner, Andrew Senior, and Koray Kavukcuoglu.
\newblock Wavenet: A generative model for raw audio.
\newblock \emph{arXiv preprint arXiv:1609.03499}, 2016.

\bibitem[Panayotov et~al.(2015)Panayotov, Chen, Povey, and
  Khudanpur]{Panayotov2015LibrispeechAA}
Vassil Panayotov, Guoguo Chen, Daniel Povey, and Sanjeev Khudanpur.
\newblock Librispeech: An asr corpus based on public domain audio books.
\newblock \emph{2015 IEEE International Conference on Acoustics, Speech and
  Signal Processing (ICASSP)}, pages 5206--5210, 2015.
\newblock URL \url{https://api.semanticscholar.org/CorpusID:2191379}.

\bibitem[Panchapagesan and Alwan(2011)]{Panchapagesan2011Chain}
Sankaran Panchapagesan and Abeer Alwan.
\newblock A study of acoustic-to-articulatory inversion of speech by
  analysis-by-synthesis using chain matrices and the maeda articulatory model.
\newblock \emph{The Journal of the Acoustical Society of America}, 129
  4:\penalty0 2144--62, 2011.
\newblock URL \url{https://api.semanticscholar.org/CorpusID:18781420}.

\bibitem[Park(2018)]{kss2018}
Kyubyong Park.
\newblock Kss dataset: Korean single speaker speech dataset, 2018.
\newblock URL
  \url{https://kaggle.com/bryanpark/korean-single-speaker-speech-dataset}.

\bibitem[Perlin(1985)]{Perlin1985}
Ken Perlin.
\newblock An image synthesizer.
\newblock \emph{SIGGRAPH Comput. Graph.}, 19\penalty0 (3):\penalty0 287–296,
  July 1985.
\newblock ISSN 0097-8930.
\newblock \doi{10.1145/325165.325247}.
\newblock URL \url{https://doi.org/10.1145/325165.325247}.

\bibitem[Radford et~al.(2022)Radford, Kim, Xu, Brockman, McLeavey, and
  Sutskever]{Radford2022Whisper}
Alec Radford, Jong~Wook Kim, Tao Xu, Greg Brockman, Christine McLeavey, and
  Ilya Sutskever.
\newblock Robust speech recognition via large-scale weak supervision.
\newblock In \emph{International Conference on Machine Learning}, 2022.
\newblock URL \url{https://api.semanticscholar.org/CorpusID:252923993}.

\bibitem[Schulze-Forster et~al.(2022)Schulze-Forster, Richard, Kelley, Doire,
  and Badeau]{SchulzeForster2022LSFDDSP}
Kilian Schulze-Forster, Ga{\"e}l Richard, Liam Kelley, Clement S.~J. Doire, and
  Roland Badeau.
\newblock Unsupervised music source separation using differentiable parametric
  source models.
\newblock \emph{IEEE/ACM Transactions on Audio, Speech, and Language
  Processing}, 31:\penalty0 1276--1289, 2022.

\bibitem[Sharma et~al.(2023)Sharma, Gero, Payne, Gruber, Rus, Torralba, and
  Andreas]{Sharma2023ContextualAC}
Pratyusha Sharma, Shane Gero, Roger Payne, David~F Gruber, Daniela Rus, Antonio
  Torralba, and Jacob Andreas.
\newblock Contextual and combinatorial structure in sperm whale vocalisations.
\newblock \emph{Nature Communications}, 15, 2023.
\newblock URL \url{https://api.semanticscholar.org/CorpusID:266150065}.

\bibitem[Shi et~al.(2015)Shi, Bu, Xu, Zhang, and Li]{AISHELL-3_2020}
Yao Shi, Hui Bu, Xin Xu, Shaoji Zhang, and Ming Li.
\newblock Aishell-3: A multi-speaker mandarin tts corpus and the baselines.
\newblock 2015.
\newblock URL \url{https://arxiv.org/abs/2010.11567}.

\bibitem[Sitzmann et~al.(2020)Sitzmann, Martel, Bergman, Lindell, and
  Wetzstein]{sitzmann2020implicit}
Vincent Sitzmann, Julien Martel, Alexander Bergman, David Lindell, and Gordon
  Wetzstein.
\newblock Implicit neural representations with periodic activation functions.
\newblock \emph{Advances in neural information processing systems (NeurIPS
  2020)}, 33:\penalty0 7462--7473, 2020.

\bibitem[Sondhi and Schroeter(1987)]{Sondhi1987}
Man Sondhi and J.~Schroeter.
\newblock A hybrid time-frequency domain articulatory speech synthesizer.
\newblock \emph{IEEE Transactions on Acoustics, Speech, and Signal Processing},
  35\penalty0 (7):\penalty0 955--967, 1987.
\newblock \doi{10.1109/TASSP.1987.1165240}.

\bibitem[Stam and Fiume(1993)]{Stam1993TurbulentWF}
Jos Stam and Eugene Fiume.
\newblock Turbulent wind fields for gaseous phenomena.
\newblock \emph{Proceedings of the 20th annual conference on Computer graphics
  and interactive techniques}, 1993.
\newblock URL \url{https://api.semanticscholar.org/CorpusID:1618202}.

\bibitem[S{\"u}dholt et~al.(2023)S{\"u}dholt, C{\'a}mara, Xu, and
  Reiss]{sudholt2023vocal}
David S{\"u}dholt, Mateo C{\'a}mara, Zhiyuan Xu, and Joshua~D Reiss.
\newblock Vocal tract area estimation by gradient descent.
\newblock 2023.

\bibitem[Takamichi et~al.(2019)Takamichi, Mitsui, Saito, Koriyama, Tanji, and
  Saruwatari]{Takamichi2019JVSCF}
Shinnosuke Takamichi, Kentaro Mitsui, Yuki Saito, Tomoki Koriyama, Naoko Tanji,
  and Hiroshi Saruwatari.
\newblock Jvs corpus: free japanese multi-speaker voice corpus.
\newblock \emph{ArXiv}, abs/1908.06248, 2019.
\newblock URL \url{https://api.semanticscholar.org/CorpusID:201070145}.

\bibitem[Tancik et~al.(2020)Tancik, Srinivasan, Mildenhall, Fridovich-Keil,
  Raghavan, Singhal, Ramamoorthi, Barron, and Ng]{tancik2020fourfeat}
Matthew Tancik, Pratul~P. Srinivasan, Ben Mildenhall, Sara Fridovich-Keil,
  Nithin Raghavan, Utkarsh Singhal, Ravi Ramamoorthi, Jonathan~T. Barron, and
  Ren Ng.
\newblock Fourier features let networks learn high frequency functions in low
  dimensional domains.
\newblock \emph{NeurIPS}, 2020.

\bibitem[Wang et~al.(2024)Wang, Sawata, Clarke, Gao, Wu, and
  Wu]{wang2024hearing}
Mason Wang, Ryosuke Sawata, Samuel Clarke, Ruohan Gao, Shangzhe Wu, and Jiajun
  Wu.
\newblock Hearing anything anywhere.
\newblock In \emph{Proceedings of the IEEE/CVF Conference on Computer Vision
  and Pattern Recognition (CVPR)}, 2024.

\bibitem[Wang et~al.(2019)Wang, Takaki, and Yamagishi]{wang2019neural}
Xin Wang, Shinji Takaki, and Junichi Yamagishi.
\newblock Neural source-filter waveform models for statistical parametric
  speech synthesis.
\newblock \emph{IEEE/ACM Transactions on Audio, Speech, and Language
  Processing}, 28:\penalty0 402--415, 2019.

\bibitem[Zhang et~al.(2018)Zhang, Isola, Efros, Shechtman, and
  Wang]{Zhang2018TheUE}
Richard Zhang, Phillip Isola, Alexei~A. Efros, Eli Shechtman, and Oliver Wang.
\newblock The unreasonable effectiveness of deep features as a perceptual
  metric.
\newblock \emph{2018 IEEE/CVF Conference on Computer Vision and Pattern
  Recognition}, pages 586--595, 2018.
\newblock URL \url{https://api.semanticscholar.org/CorpusID:4766599}.

\end{thebibliography}
\bibliographystyle{plainnat}
\newpage
\appendix
\section{Physics of the Vocal Tract}
We understand that the physics of speech production may be unfamiliar to the broader machine learning community, so we provide a derivation of the speech simulation method in this section. For a comprehensive introduction to speech, we refer the reader to~\citet{Flanagan1972}, or~\citet{BilbaoNumerical2009} for a more general overview on physical modeling synthesis. 
\label{app:webster_derivation}
\subsection{Derivation of the Linearized Euler Equations}
 In the most general form, we can model the fluid dynamics of the vocal tract with Euler's equations for compressible flow, over 3D space $\mathbf{x} \in \mathbb{R}^3$ and time $t \in \mathbb{R}$. For a pressure field $p:(\mathbf{x}, t)  \mapsto \mathbb{R}$, velocity field $\mathbf{v}: (\mathbf{x}, t) \mapsto \mathbb{R}^3$, and density field $\rho: (\mathbf{x}, t) \mapsto \mathbb{R}_{\geq 0}$, the first equation represents conservation of mass, while the second represents conservation of momentum. Subscripts $_t$ and $_x$ denote partial derivatives in time and space respectively. 
\begin{align}
    \rho_t &= - \nabla \cdot (\rho \mathbf{v}) &&\text{mass}\\
     - \nabla p  &= \rho  ({\mathbf{v}}_t + ({\mathbf{v}} \cdot \nabla) \mathbf{v})&&\text{momentum}
\end{align}
The pressure at the outlet of the vocal tract, which we will denote with $p_{\text{out}}$, is what we perceive as speech. These equations are difficult to model numerically, but by applying a series of simplifying assumptions, we can derive an entire hierarchy of governing equations for vocal tract acoustics. 

First, when modeling sound propagation, $p$ and $\rho$ may be defined as perturbations around equilibrium states: $\rho = \rho_0  + \rho'$, $p = p_0 + p'$. Substituting these linearizations into Euler's equations, we have
\begin{align}
    (\rho_0 + \rho')_t &= -\nabla \cdot  ((\rho_0 + \rho')\mathbf{v})\\
    - \nabla(p_0 + p') &= (\rho_0 + \rho') ({\mathbf{v}}_t + ({\mathbf{v}} \cdot \nabla) \mathbf{v}).
\end{align}
Because $v$, $\rho'$ and $p'$ are assumed to be small, all second-order terms may be eliminated. Further, we may assume that $p' = c^2 \rho'$, where $c$ is the speed of sound, eliminating the need for $\rho'$. The resulting equations are a common form of Euler's equations used in acoustics. Note that by linearization, we lose the ability to synthesize turbulent flow. 
\begin{align}
     \frac{1}{c^2}p'_t &= - \rho_0 \nabla \cdot \mathbf{v} && \text{mass}\label{eq:euler1}\\
    -\nabla{p'} &= \rho_0 \mathbf{v}_t &&\text{momentum}\label{eq:euler2}
\end{align}
For frequencies less than 4 kHz, the acoustics of the vocal tract are dominated by plane waves: waves traveling perpendicular to the vocal tract's cross section $\rm{d}A = A(x)~\rm{d}\mathbf{\hat{n}}$. This is because the wavelength of a 4 kHz wave traveling at the speed of sound (343 m/s) is $8.6$ cm, significantly larger than the diameter of the vocal tract (2-3 cm). This is a common simplification used in prior works~\citep{kelly_lochbaum_1962,Maeda1982,Sondhi1987,BirkholzVTL}. We can model planar flow by rewriting Eq.~\ref{eq:euler1} and Eq.~\ref{eq:euler2} in terms of a one-dimensional flux, $u(x, t) = \int_{A(x)} \mathbf{v} \cdot \rm{d}A$. To model this flux  (also known as volume velocity), we integrate the governing equations over a control volume $V$. 
\begin{align}
\int_V\frac{1}{c^2}p'_t~\rm{d}V&= - \rho_0  \int_{V}(\nabla \cdot \mathbf{v})~\rm{d}V\\
    -\int_{V}\nabla{p'}~ \rm{d}V&= \rho_0 \int_{V}\mathbf{v}_t~ \rm{d}V.
\end{align}
In the limit, $\rm{d}V = \rm{d}A~\rm{d}x$. Further assuming that $p'(\mathbf{x}, t)$ is constant over each $\rm{d}A$, the 3D field $p'(\mathbf{x}, t)$ can be replaced with a simpler 1D field: $ p'(x, t)$.
By applying the divergence theorem and expanding the integrals, we find that
\begin{align}
A\Delta x\frac{1}{c^2}p'_t&= - \rho_0  \int_{A(x) + \Delta x}\mathbf{v} \cdot ~\rm{d}A - \rho_0 \int_{A(x) }\mathbf{v} \cdot ~\rm{d}A\\
    -A\Delta x ~p'_x&= \rho_0 \int_{x}^{x + \Delta x} \int_A\mathbf{v}_t~\cdot \rm{d}A~\rm{d}x.
\end{align}
Finally, substituting $u$, and taking the limit $\Delta x \mapsto 0$, we derive Euler's equations in terms of $u$ and $A$. For simplicity, $p'$ is often written as just $p$, and $\rho_0$ is written as $\rho$, the ambient density of air. 
\begin{align}
\frac{A}{c^2}p_t&= - \rho  u_x&&\text{mass}\label{eq:lee-mass}\\
    -A ~p_x&= \rho u_t&&\text{momentum}\label{eq:lee-momentum}
\end{align}

These are the basic equations for modeling acoustics in the vocal tract, and can be easily augmented with additional dampening parameters for the vocal tract's material properties~\citep{Flanagan1972}. 
The equations can be written even more compactly by differentiating the mass equation by $t$, the momentum equation by $x$, then combining them a single equation, known as Webster's equation:
\begin{align}
    (A(x)p_x)_x = & \frac{A(x)}{c^2} p_{tt} 
\end{align}
Importantly, Webster's equation resembles the wave equation, demonstrating how the geometry of the vocal tract, controlled by $A(x)$, affects the wave propagation of speech. After applying the frequency domain separation of variables $p(x, t) = P(x)e^{-j\omega t}$, $u(x, t) = U(x)e^{-j\omega t}$, Webster's equation becomes
\begin{align}
(A(x)P(x)_x)_x = -\frac{\omega^2}{c^2}A(x)P(x),
\end{align}
resembling the Helmholtz equation. Solving the linearized Euler equations or Webster's equation is thus equivalent to solving the Helmholtz eigenvalue problem in the frequency domain. 

To generate vowels, a glottal pulse $u_{\text{in}}$ parameterized by the LF model~\citep{fant1985four} is set as the initial condition, and is propagated to the other end of the tract to synthesize $p_{\text{out}}$. Solving the linearized Euler equations or Webster's equation derives a filter $h_{\text{tract}}$, such that $p_{\text{out}} = h_{\text{tract}}* u_{\text{in}}$.

The following boundary conditions at the glottis and the lip are typically adopted to solve the system.
\begin{align}
    u(0, t) &= u_{\text{in}}(t)  & \text{Glottis boundary}\label{eq:bc1} \\
    \quad R_\text{lips} p(L, t) + L_\text{lips} p_t (L, t) &= R_{\text{lips}}L_\text{lips}  u_t(L, t) & \text{Lip boundary}\label{eq:bc2}
\end{align}
The lip opening area, $A(L)$, controls how much sound is reflected back into the tract versus how much is dissipated outside via  an inductance term $L_\text{lips} = \frac{8\rho}{3\pi \sqrt{\pi A(L)}}$ and resistance term $R_\text{lips} = \frac{128\rho c}{9\pi^2 A(L)}$ (with coefficients derived by~\citet{Flanagan1972}).

\subsection{Circuit Interpretation and the Transmission Line Model}
\label{subsec:circuit_analogy}
Interpreting the fluid dynamics of the vocal tract as a circuit model has been long studied \cite{dunn1950calculation}.
Considering a piecewise-constant area function with $N$ cylindrical sections where each $n$-th section has of length $l_n$ and area $A_n$ and taking the Fourier transform of Eq.~\eqref{eq:lee-mass}-\eqref{eq:lee-momentum} gives the following frequency-domain governing equations:
\begin{align}
    j\omega C_n P_n(\omega) &= U_n(\omega) - U_{n+1}(\omega), \\
    j\omega L_n U_n(\omega) &= P_{n-1}(\omega) - P_n(\omega),
\end{align}
where $C_n=A_n l_n/(\rho_0 c^2)$ is the acoustic \emph{compliance} and $L_n=\rho_0 l_n /(2A_n)$ is the acoustic \emph{inductance} of the half-section.
This describes a lossless ideal fluid flow with purely imaginary impedances.
\paragraph{Extension with viscous losses.}
More realistic fluid result from a viscous boundary layer at the cylinder wall.
This can be modeled by imposing the no-slip condition at the wall, so the resulting wall shear stress acts as a drag on the fluid.
This `wall friction' due to this boundary layer is known to be well-modeled by introducing an additional dissipative term in the momentum equation:
\begin{equation}\label{eq:lee-momentum-wall-friction}
    -\nabla p = L_n \dot u + r_{\mathrm{wf}}(t) * u
\end{equation}
where $*$ denotes convolution and $r_\mathrm{wf}(t)$ is the impulse response of the wall-friction operator. For a circular cylinder of perimeter $S_n$ and cross-sectional area $A_n$, the viscous boundary-layer theory gives the frequency-domain expression of this wall-friction operator as:
\begin{equation}\label{eq:resistance}
    R_n(x,\omega) = \frac{S_n(x)}{2A_n^2(x)} \sqrt{\frac{\rho_0\omega\mu}{2}}
\end{equation}
which can be analogously interpreted as a frequency-dependent \emph{resistance} $R_n(\omega)\propto\sqrt{\omega}$. 
Taking the Fourier transform to Eq.~\eqref{eq:lee-momentum-wall-friction} gives the follows.
\[P_{n-1}(\omega) - P_n(\omega) = (j\omega L_n + R_n(\omega)) U_n(\omega)\]
The series impedance of the cylinder at the $n$-th section is therefore
\begin{equation}\label{eq:z-series}
    Z_{s,n}(\omega) = R_n(\omega) + j\omega L_n.
\end{equation}

\paragraph{Extension with yielding walls.}
In addition to modeling the fluid flow, the acoustic modeling of the vocal tract wall can be approached using a damped harmonic oscillator (equivalently the mass-spring-damper).
Denoting the wall mass, damping coefficient, and stiffness coefficient as $m_w$, $b_w$, and $k_w$, respectively, the normal displacement of the wall $\xi_n$ satisfies
\begin{equation}\label{eq:wall-dho}
    m_w\ddot\xi_n + b_w\dot\xi_n + k_w\xi_n = p(x,t).
\end{equation}
In the frequency domain, Eq.~\eqref{eq:wall-dho} gives a wall displacement $\Xi_n=P/(-\omega^2 m_w + j\omega b_w + k_w)$, so the wall presents a shunt impedance to the acoustic field.
For $n$-th section with lateral surface area $S_nl_n$, the lumped wall impedance is
\[Z_{w,n} = \frac{-\omega^2 m_w + j\omega b_w + k_w}{S_nl_n\omega^2} = \frac{R_{w,n} + j\omega L_{w,n} + \tfrac{1}{j\omega C_{w,n}}}{j\omega}\]
Combined with the acoustic compliance $C_n$ in parallel, the total shunt admittance is
\begin{equation}\label{eq:z-wall}
    Z_{w,n}(\omega) = \left(R_{w,n} + j\omega L_{w,n} + \frac{1}{j\omega C_{w,n}}\right) \parallel \frac{1}{j\omega C_n}
\end{equation}
where $L_{w,n}=m_w/(S_nl_n)$, $R_{w,n}=b_w/(S_nl_n)$, and $C_{w,n}=(S_nl_n)/k_w$.

\subsection{Transfer Functions}
\label{subsec:transfer_function}
Combining Eq.~\eqref{eq:z-series} and Eq.~\eqref{eq:z-wall}, each $n$-th section of the acoustic tube is represented in the frequency domain as the \emph{T-network} where a series arm impedance $Z_{s,n}$ carries momentum losses and a shunt arm $Z_{w,n}$ stores compliance and wall losses.
Finally, as described in Eq.~\eqref{eq:transfer-function}, the two-port transfer matrix for section $n$ is
\begin{equation}
  \begin{bmatrix} P_{n+1} \\ U_{n+1} \end{bmatrix}
  = \underbrace{\begin{bmatrix}
      1 + Z_{s,n}/Z_{w,n} & -2Z_{s,n} - Z_{s,n}^2/Z_{w,n}\\
      -1/Z_{w,n} & 1 + Z_{s,n}/Z_{w,n}
    \end{bmatrix}}_{\mathbf{K}_n(\omega)}
  \begin{bmatrix} P_n \\ U_n \end{bmatrix}
  \label{eq:twoport}
\end{equation}
with a section matrix
\[\mathbf{K}_n = \begin{bmatrix}\mathbf{A}_n & \mathbf{B}_n \\ \mathbf{C}_n & \mathbf{D}_n\end{bmatrix}.\]
This can be used to construct partial chain matrices by concatenating the individual section matrices:
\begin{align}
    \mathbf{K}_{0\mapsto n} &:= \mathbf{K}_n\mathbf{K}_{n-1}\cdots \mathbf{K}_1 = \begin{bmatrix}
        \mathbf{A}_{0n} & \mathbf{B}_{0n} \\ \mathbf{C}_{0n} & \mathbf{D}_{0n}
    \end{bmatrix}, \\
    \mathbf{K}_{n\mapsto N+1} &:= \mathbf{K}_N\cdots \mathbf{K}_{n+1} = \begin{bmatrix}
        \mathbf{A}_{nN} & \mathbf{B}_{nN} \\ \mathbf{C}_{nN} & \mathbf{D}_{nN}
    \end{bmatrix}.
\end{align}
Note that chain matrices have unit determinant for reciprocal media, \textit{i.e.}, $\det\mathbf{K}_{0\mapsto n} =\det\mathbf{K}_{n\mapsto N+1}=1$.
\paragraph{Transfer function from glottis to section $n$.}
To compute $H_{0\mapsto n}$, the transfer function from glottis to section $n$, apply Eq.~\eqref{eq:twoport} for $\mathbf{K}_{0\mapsto n}$:
\begin{equation}
  \begin{bmatrix} P_{n} \\ U_{n} \end{bmatrix}
  = \begin{bmatrix}
      \mathbf{A}_{0n} & \mathbf{B}_{0n} \\ \mathbf{C}_{0n} & \mathbf{D}_{0n}
    \end{bmatrix}
  \begin{bmatrix} P_\mathrm{in} \\ U_\mathrm{in} \end{bmatrix}
  \label{eq:twoport-0-n}
\end{equation}
where $P_\mathrm{in}$ is the pressure at the glottal plane and $U_\mathrm{in}$ is the glottal volume velocity.
Imposing the boundary condition at the glottis, $P_\mathrm{in} = Z_\mathrm{in} U_\mathrm{in}$,
\[U_n = (\mathbf{C}_{0n} Z_\mathrm{in} + \mathbf{D}_{0n})U_\mathrm{in},\]
\begin{equation}
    G_{0\mapsto n}(\omega) = \frac{U_n}{U_\mathrm{in}} = \mathbf{C}_{0n} Z_\mathrm{in} + \mathbf{D}_{0n}.
\end{equation}
The backward impedance seen from section $n$ looking toward the glottis follows from the same substitution.
\[Z_1 = -\frac{P_n}{U_n} = -\frac{\mathbf{A}_{0n} Z_\mathrm{in} + \mathbf{B}_{0n}}{\mathbf{C}_{0n} Z_\mathrm{in} + \mathbf{D}_{0n}}\]
The sign convention for $-U_n$ indicates that $U_n$ is flowing away from glottis and thus into the $Z_1$ load when viewed from section $n$.

\paragraph{Transfer function from section $n$ to lips.}
From boundary condition Eq.~\eqref{eq:bc2}, the radiation impedance $Z_\mathrm{lips} = R_\mathrm{lips} \,\parallel\, j\omega L_\mathrm{lips}$ can be derived with
\begin{equation}
  L_\mathrm{lips} = \frac{8\rho_0}{3\pi\sqrt{\pi A_N}} \qquad\mathrm{and}\qquad
  R_\mathrm{lips} = \frac{128\rho_0 c}{9\pi^2 A_N}.
  \label{eq:Zrad}
\end{equation}
Now, apply $\mathbf{K}_{n\mapsto N+1}$ from section $n$ to the lip plane, where the boundary condition $P_{N+1} = Z_\mathrm{lips}U_{N+1}$ must be satisfied:
\begin{equation}
  \begin{bmatrix} P_{N+1} \\ U_{N+1} \end{bmatrix}
  = \begin{bmatrix}
        \mathbf{A}_{nN} & \mathbf{B}_{nN} \\ \mathbf{C}_{nN} & \mathbf{D}_{nN}
    \end{bmatrix}
  \begin{bmatrix} P_n \\ U_n \end{bmatrix},
  \label{eq:twoport-n-N+1}
\end{equation}
giving
\begin{align}
    P_{N+1} &= \mathbf{A}_{nN} P_n + \mathbf{B}_{nN} U_n, \label{eq:tf-pout-linear-expand} \\
    U_{N+1} &= \mathbf{C}_{nN} P_n + \mathbf{D}_{nN} U_n. \label{eq:tf-uout-linear-expand}
\end{align}
Substituting $P_{N+1} = Z_\mathrm{lips}U_{N+1}$ into Eq.~\eqref{eq:tf-pout-linear-expand} and \eqref{eq:tf-uout-linear-expand} to eliminate $P_{N+1}$:
\begin{equation}\label{eq:tf-intermediate}
    Z_\mathrm{lips} U_{N+1} = \mathbf{A}_{nN} P_n + \mathbf{B}_{nN} U_n.
\end{equation}
Divide Eq.~\eqref{eq:tf-intermediate} by \eqref{eq:tf-uout-linear-expand}:
\begin{equation}
    Z_\mathrm{lips} = \frac{\mathbf{A}_{nN} P_n + \mathbf{B}_{nN} U_n}{\mathbf{C}_{nN} P_n + \mathbf{D}_{nN} U_n}
    \qquad\Longrightarrow\qquad
    Z_2:=\frac{P_n}{U_n} = \frac{\mathbf{D}_{nN} Z_\mathrm{lips} - \mathbf{B}_{nN}}{\mathbf{A}_{nN} - \mathbf{C}_{nN}Z_\mathrm{lips}}
\end{equation}
which is the forward impedance seen from section $n$ towards the lips. Eliminating $P_n$ using $P_n=Z_2U_n$ yields:
\[U_{N+1} = \left(\mathbf{C}_{nN} Z_2 + \mathbf{D}_{nN}\right) U_n = \frac{\mathbf{C}_{nN}(\mathbf{D}_{nN} Z_\mathrm{lips} - \mathbf{B}_{nN}) + \mathbf{D}_{nN}(\mathbf{A}_{nN} - \mathbf{C}_{nN}Z_\mathrm{lips})}{\mathbf{A}_{nN} - \mathbf{C}_{nN}Z_\mathrm{lips}} U_n.\]
Because $\mathbf{A}_{nN}\mathbf{D}_{nN}
-
\mathbf{B}_{nN}\mathbf{C}_{nN}
=\det\mathbf{K}_{n\mapsto N+1}=1$, the numerator simplifies, leaving
\begin{equation}
    G_{n\mapsto N+1} = \frac{U_{N+1}}{U_n} = \frac{1}{\mathbf{A}_{nN} - \mathbf{C}_{nN}Z_\mathrm{lips}}.
\end{equation}
The pressure transfer function from $n$ to lips is therefore
\begin{equation}
    H_{n\mapsto N+1} = \frac{P_\mathrm{out}}{U_n} = \frac{Z_\mathrm{lips}}{\mathbf{A}_{nN} - \mathbf{C}_{nN}Z_\mathrm{lips}}.
\end{equation}

\paragraph{Combining the full and noise transfer functions.}
The glottis-to-lips volume velocity transfer is obtained by applying $\mathbf{K}_\mathrm{tot} = \mathbf{K}_N \mathbf{K}_{N-1}\cdots \mathbf{K}_1$ as
\begin{equation}
  \begin{bmatrix} P_{N+1} \\ U_{N+1} \end{bmatrix}
  = \begin{bmatrix}
      \mathbf{A}_{0N} & \mathbf{B}_{0N} \\ \mathbf{C}_{0N} & \mathbf{D}_{0N}
    \end{bmatrix}
  \begin{bmatrix} P_\mathrm{in} \\ U_\mathrm{in} \end{bmatrix}.
  \label{eq:twoport-0-N+1}
\end{equation}
Setting $P_\mathrm{in}=Z_\mathrm{in}U_\mathrm{in}$ at the glottis and imposing $P_{N+1}=Z_\mathrm{lips}U_{N+1}$ at the lips gives
\[\mathbf{A}_{0N} Z_\mathrm{in} + \mathbf{B}_{0N} = Z_\mathrm{lips}(\mathbf{C}_{0N} Z_\mathrm{in} + \mathbf{D}_{0N})\]
therefore the input impedance of the tract seen from the glottis is
\begin{equation}
    Z_\mathrm{in} = \frac{\mathbf{D}_{0N}Z_\mathrm{lips} - \mathbf{B}_{0N}}{\mathbf{A}_{0N} - \mathbf{C}_{0N}Z_\mathrm{lips}}.
    \label{eq:input-impedance}
\end{equation}
With $Z_\mathrm{in}$ established, substitute Eq.~\eqref{eq:input-impedance} into $H_\mathrm{tract} = P_{N+1}/U_\mathrm{in}=\mathbf{A}_{0N} Z_\mathrm{in} + \mathbf{B}_{0N}$ gives:
\begin{align}
    H_\mathrm{tract}(\omega) &= \mathbf{A}_{0N} \left(\frac{\mathbf{D}_{0N}Z_\mathrm{lips} - \mathbf{B}_{0N}}{\mathbf{A}_{0N} - \mathbf{C}_{0N}Z_\mathrm{lips}}\right) + \mathbf{B}_{0N}, \\
    &= \frac{\mathbf{A}_{0N}(\mathbf{D}_{0N}Z_\mathrm{lips} - \mathbf{B}_{0N} ) + \mathbf{B}_{0N}(\mathbf{A}_{0N} - \mathbf{C}_{0N} Z_\mathrm{lips})}{\mathbf{A}_{0N} - \mathbf{C}_{0N} Z_\mathrm{lips}}.
\end{align}
The numerator expands to $(\mathbf{A}_{0N}\mathbf{D}_{0N} - \mathbf{B}_{0N}\mathbf{C}_{0N})Z_\mathrm{lips} = \det \mathbf{K}_\mathrm{tot}\cdot Z_\mathrm{lips} = Z_\mathrm{lips}$. Therefore, the total pressure transfer function is
\begin{equation}
    H_\mathrm{tract} = \frac{Z_\mathrm{lips}}{\mathbf{A}_{nN} - \mathbf{C}_{nN}Z_\mathrm{lips}}
\end{equation}

As described in Subsection \ref{subsec:consonants}, the propagation of the volume-velocity noise source $U_{\text{noise}_n}(\omega)$ to the lips is is only governed by $\mathbf{K}_{n\mapsto N+1}$, and therefore the pressure transfer function for the noise from the section $n$ is identical to $H_{n\mapsto N+1}$.
This confirms that the total combined transfer function from the glottal pulse and the noise to the lips as Eq.~\eqref{eq:tf-tract-plus-noise}.

\subsection{The Nasal Tract}
Although we do not pursue nasal tract reconstruction because the nasal tract does not appear in the MRI frames, the transmission line model can be used to model it. 
In fact, with the TLM method, the nasal effect can be modeled as a bifurcated transmission line, and the combination of these two LTI systems can be represented as an equivalent circuit consisting of a single transmission line.
In the present work as well, although the nasal cavity is not explicitly reconstructed, it can be viewed as being modeled using a lumped equivalent area function. Examples of this technique are provided in~\citet{Sondhi1987}.

\section{Finite Difference Techniques}
\label{app:finite_differences}

\subsection{Time Domain Governing Equation}\label{sec:time-domain-ge}
Solving Eq.~\eqref{eq:lee-mass} with the wall-friction momentum equation Eq.~\eqref{eq:lee-momentum-wall-friction} requires a convolution with a half-order integro-differential operator for frequency-dependent resistance $R(\omega)\propto\sqrt{\omega}$. In practice, the resistance is implemented using Hagen--Poiseuille DC resistance \cite{Maeda1982}
\[R_n^\mathrm{DC} = \frac{4\mu l_n\pi}{A_n^2}\]
which is the Stokes flow resistance for a circular duct in the zero-frequency limit.
With this substitution, all terms in the circuit equations for the $N$-section vocal tract model reduce to the following coupled ODEs.
\begin{align}
    \dot p_n &= \frac{1}{C_n}(u_n - u_{n+1} - u_{w,n}) &\text{mass}\label{eq:time-continuous-mass} \\
    (L_{n-1} + L_n)\dot u_n + (R_{n-1} + R_n) u_n &= p_{n-1} - p_n & \text{momentum} \\
    L_{w,n}\ddot u_{w,n} + R_{w,n}\dot u_{w,n} + \frac{u_{w,n}}{C_{w,n}} &= \dot p_n & \text{wall vibration}
\end{align}
The expression for the wall vibration can be deduced from Eq.~\eqref{eq:wall-dho} by taking time derivative and denoting $u_{w,n}=S_{n}l_n\dot\xi_n$.
The last section $i=N$ terminates into the radiation impedance $Z_\mathrm{lips}=R_\mathrm{lips}\parallel j\omega L_\mathrm{lips}$. In the time domain this parallel $R$--$L$ load introduces two additional flow unknowns $u_{N+1}$ (total lip flow) and $u_{N+2}$ (inductive branch flow):
\begin{align}
  \dot u_{N+1} L_N + u_{N+1}(R_N + R_\mathrm{lips}) - u_{N+2} R_\mathrm{lips} &= p_N, \label{eq:total-lip-flow}\\
  \dot u_{N+2} L_\mathrm{lips} + u_{N+2} R_\mathrm{lips} - u_{N+1} R_\mathrm{lips} &= 0. \label{eq:inductive-branch-flow}
\end{align}
The radiated pressure (voltage across $R_\mathrm{lips}$) is
\begin{equation}
  p_\mathrm{out}(t) = R_\mathrm{lips}\bigl(u_{N+1}(t) - u_{N+2}(t)\bigr)
                    = L_\mathrm{lips}\,\dot u_{N+2}(t).
  \label{eq:prad-cont}
\end{equation}
These equations describe the governing equations in continuous time-domain.
The only departure from the frequency-domain expressions is the substitution of $R_n(\omega)$ by $R_n^\mathrm{DC}$.

\subsection{Numerical Scheme}

In order to solve the time-domain governing equation on a discrete stencil,  the semi-implicit scheme has been employed:
\begin{equation}
  \dot f = \frac{f - f'}{\Delta t\,\vartheta} - \frac{\bar\vartheta}{\vartheta}\,\dot f',
  \qquad \bar\vartheta = 1-\vartheta.
  \label{eq:theta-scheme}
\end{equation}
While various schemes may result depending on the value of $\vartheta$, in practice, we set $\vartheta = 0.5$ (trapezoidal) to perform time-domain simulations.
The process of solving Subsection \ref{sec:time-domain-ge} with this scheme reduces to a tridiagonal linear system, which can be resolved through the time-domain recursion and \texttt{jax.linalg.tridiagonal\_solve}.
Please note that in this study, this time-domain simulation serves only as a baseline for validating the performance of the employed frequency-domain simulator.
For further details regarding the time-domain implementation, please refer to \citet{Birkholz2004InfluenceOT}.

\section{Experimental Details}
\label{app:experimental_details}
\begin{figure}[t]
\centering
\begin{subfigure}{0.28\textwidth}
    \centering
    \includegraphics[width=\textwidth]{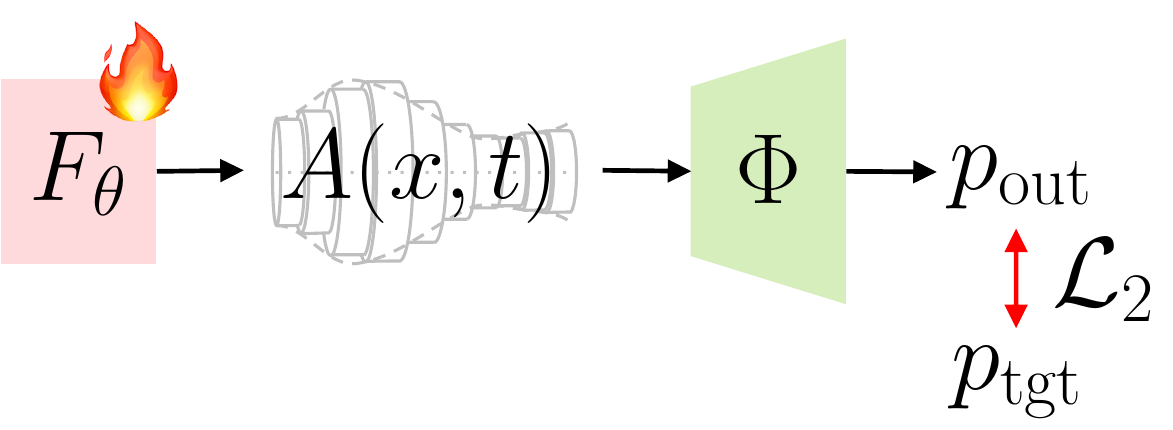}
    \caption{Neural field training}
    \label{fig:schematics-inr}
\end{subfigure}
\hfill
\begin{subfigure}{0.3\textwidth}
    \centering
    \includegraphics[width=\textwidth]{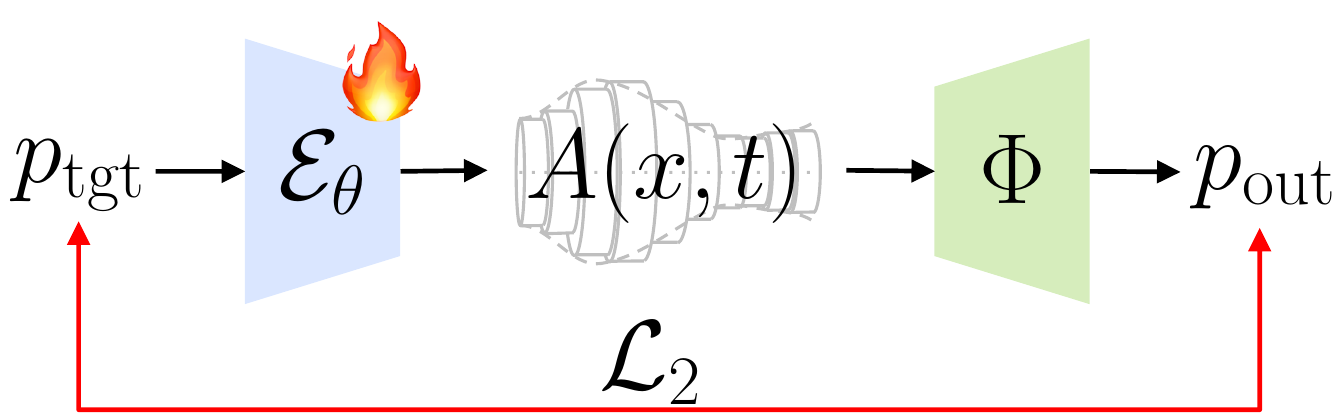}
    \caption{Self-supervised autoencoding}
    \label{fig:schematics-ae}
\end{subfigure}
\hfill
\begin{subfigure}{0.36\textwidth}
    \centering
    \includegraphics[width=\textwidth]{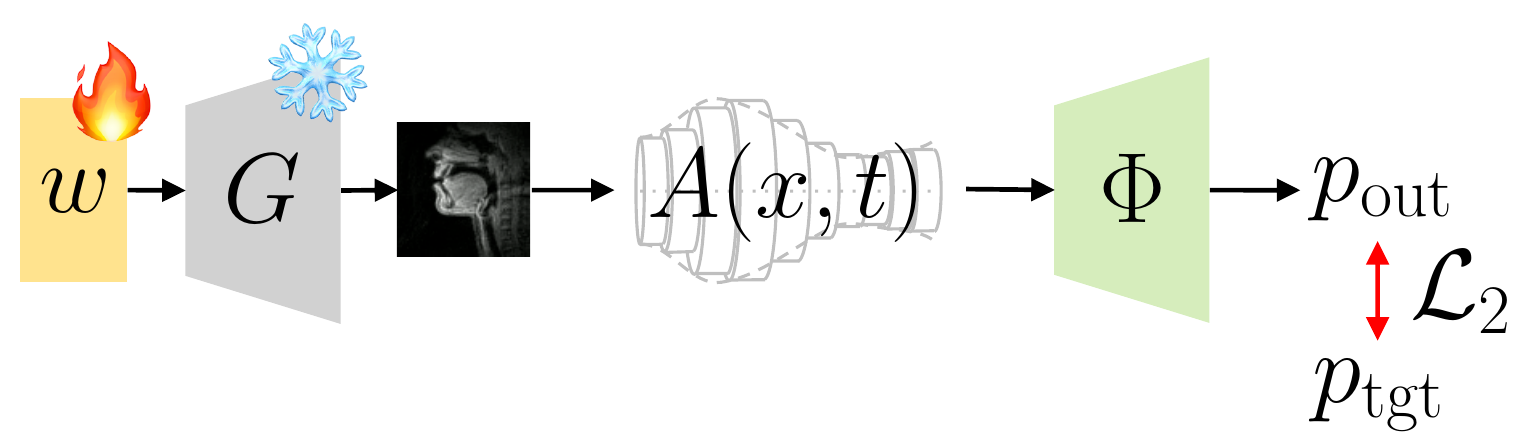}
    \caption{GAN latent optimization}
    \label{fig:schematics-gan}
\end{subfigure}
\caption{
    A schematic diagram illustrating three application cases of the differentiable $\Phi$, as described in Section \ref{sec:vocal_recon}.
    The problem of estimating $A(x,t)$ from speech can thus be solved through various gradient-based optimization techniques, and \textit{none} of these methods presuppose a ground truth $A$, but rather resolve this inverse problem solely by utilizing the real-world speech signal $p_\mathrm{tgt}$.
}
\label{fig:schematics}
\end{figure}

\subsection{Simulation Details}
The simulation is modeled in centimeters. For all experiments, we use $N=32$ spatial segments. For the benchmarks in the paper, and in general, we run our simulation at 16 kHz. Higher fidelity can however be achieved with higher sample rates.  For the singing examples in the supplementary material, we use sample rates of 32 kHz - 44.1 kHz.

\textbf{Glottal Pulse.} To estimate the glottal pulse $u_\text{in}$ of each signal, we first extract fundamental frequencies from the speech sample using Swift-F0~\citep{nieradzik2025swiftf0}. The fundamental frequencies are then used to construct a glottal pulse using the LF model~\citep{fant1985four}. Its time-varying amplitude is determined by the RMS energy of the signal, measured in Librosa. The pulse is normalized so it has a maximum value of 1 $\rm{cm}^3/\rm{s}$. Generally, we find that gradient-based optimization is insensitive to the glottal pulse used. The spectrogram loss focuses gradients on formants, which are a property of the vocal tract's resonances and not the glottal pulse. To show robustness to glottal pulse, we include a vocal tract reconstruction of a song played on a piano in the supplementary material. The model is able to create an acapella reconstruction of the song, Ode to Joy, while maintaining a human-like timbre. 

\subsection{Loss Function}\label{subsec:loss}
As described in Section \ref{sec:problem_statement}, we use a multiscale log Mel spectrogram loss to fit the speech samples. We use $n_\mathrm{mel}=128$ Mel bins, FFT sizes of $(512, 1024, 2048)$, window lengths of $(160, 400, 800)$, and hop lengths of $(40, 80, 160)$.
This builds three configurations for the spectrogram resolution: $\mathcal{C}=\{(512, 160, 40), (1024, 400, 80), (2048, 800, 160)\}$.
For a target pressure $p_\mathrm{tgt}$ and an output pressure $p_{\text{out}} \triangleq \Phi(u_{\text{in}}, A_\theta)$, the loss is computed using the $L_2$ norm of the log Mel spectrogram distances under three configurations $(n_\mathrm{fft}, n_\mathrm{win}, n_\mathrm{hop})\in\mathcal{C}$ averaged over the choices
\begin{equation}\label{eq:loss}
    \mathcal{L}_2(p_{\text{out}},p_{\text{tgt}})
    = \frac{1}{|\mathcal{C}|}\sum_{\mathcal{C}} \left\lVert \log \text{mel}(p_{\text{out}}) - \log \text{mel}(p_{\text{tgt}}) \right\rVert_2.
\end{equation}
Here, $\log\text{mel}$ is the log Mel spectrogram transformation using the Mel filterbank $\mathbf{M}\in\mathbb{R}^{n_\mathrm{mel}\times n_\mathrm{freq}}$ and the short-time Fourier transformation (STFT) that outputs the spectrogram $\mathbf{p}\in\mathbb{R}^{n_\mathrm{freq}\times K}$, where $n_\mathrm{freq}=\lfloor n_\mathrm{fft}/2\rfloor+1$ is the number of STFT frequency bins and $K=\lceil(n_\mathrm{sample}-n_\mathrm{window})/n_\mathrm{hop}\rceil+1$ is the number of its window chunks.
The $m$-th Mel bin of the $k$-th window of the log Mel spectrogram is computed as $\log\text{mel}(p)[m,k]=\log(\sum_j \mathbf{M}[m,j]\cdot|\mathbf{p}[j,k]|)$.
It is well-known that this trick of employing a multi-resolution spectrogram can distribute the resolution dependency of the loss calculation, thereby facilitating smoother optimization.

\subsection{Evaluation Setup for Perceptual Speech Quality Metrics}
\label{app:experimental_setup}
We evaluate how well each method can fit 25 random speech samples from LibriTTS-R~\citep{Koizumi2023LibriTTSRAR}. For the results in Table~\ref{tab:perceptibility}, we parametrize each area function with a multiplicative filter network~\citep{Fathony2021MultiplicativeFN} (this choice is explained in Subsection~\ref{subsec:neural_fields}). Gradient descent is done with the Adam optimizer~\citep{Kingma2014AdamAM} with a learning rate of $10^{-2}$ for 200 steps.

Using TorchAudio-Squim~\citep{Kumar2023TorchaudioSquimRS}, we report SI-SDR (scale-invariant signal-to-distortion ratio), STOI (short-time objective intelligibility), and PESQ (wideband perceptual evaluation of speech quality). 

\subsection{Neural Fields}
We study two neural field parameterizations $F_\theta$ in the paper, a 4-layer fully connected network with random Fourier feature (RFF) positional encoding~\citep{tancik2020fourfeat}, and a 3-layer multiplicative filter network (MFN)~\citep{Fathony2021MultiplicativeFN}. The RFF network has a 64-dimensional encoding, and a 256-dimensional hidden dimension. The MFN network had an input scale of 256, and 256-dimenaional hidden dimension. To constrain the area functions to be positive, we apply a softmax to the output, and clip the minimum values at $10^{-2}$. All models were trained with the Adam optimizer~\citep{Kingma2014AdamAM} for 200 steps and a learning rate of $10^{-2}$. To fit a 5 second speech sample, training takes approximately 1 minute.
\begin{wrapfigure}{U}{.43\textwidth}
     \centering
     \vspace{-10pt}
        \captionof{table}{\textbf{Neural Field Reconstruction}}
        \scalebox{0.76}{
        \begin{tabular}{l|ccc}
        \toprule
        &\textbf{Discrete} & \textbf{RFF}~\citep{tancik2020fourfeat} & \textbf{MFN}~\citep{Fathony2021MultiplicativeFN} \\
        \midrule
        SI-SDR & 10.78 {\scriptsize $\pm$2.22} & 14.51 {\scriptsize $\pm$3.78}& \textbf{16.39} {\scriptsize $\pm$2.42} \\ 
        STOI & 0.81 {\scriptsize $\pm$0.02}& 0.84 {\scriptsize $\pm$0.05}& \textbf{0.93} {\scriptsize $\pm$0.02}\\
        PESQ & 1.57 {\scriptsize $\pm$0.12}& 1.76 {\scriptsize $\pm$0.26}& \textbf{1.98} {\scriptsize $\pm$0.26}\\

        \bottomrule
        \end{tabular}
        }
        \vspace{-10pt}
    \label{tab:field_comparison}
\end{wrapfigure}

\textbf{Neural Fields Improve Acoustic Reconstruction.}  
We also study how each field performs at reconstructing real-world speech in Table~\ref{tab:field_comparison}. Using the experimental setup in Appendix~\ref{app:experimental_setup}, we reconstruct the area functions for 25 speech samples from~\citep{Koizumi2023LibriTTSRAR}. We find that both neural fields significantly outperform the discrete case across all speech metrics. 

\subsection{Autoencoding}
\label{app:autoencoding}
We use the Wav2Vec 2.0~\citep{Baevski2020wav2vec2A} architecture for the autoencoder. Raw waveforms are input into a CNN with the same dimension as the original Wav2Vec2 model. The CNN embeddings are then passed into a 12 layer transformer with 12 heads, and a hidden dimension of 768. The model is trained with the Adam optimizer~\citep{Kingma2014AdamAM} at a learning rate of $10^{-4}$. The model is trained on 3 second segments of English and a batch size of 64 for 15,000 iterations, equating to 800 hours of training audio. It is then fine-tuned for each other language for 10 epochs.

We train our encoder on LibriTTS-R~\citep{Koizumi2023LibriTTSRAR} for English, Multilingual LibriSpeech~\citep{Panayotov2015LibrispeechAA} for other European languages, KSS~\citep{kss2018} for Korean, JVS~\citep{Takamichi2019JVSCF} for Japanese, and AISHELL-3~\citep{AISHELL-3_2020} for Chinese.
\subsubsection{UMAP Projection}
\begin{figure}[t]
\centering
\includegraphics[width=1\linewidth]{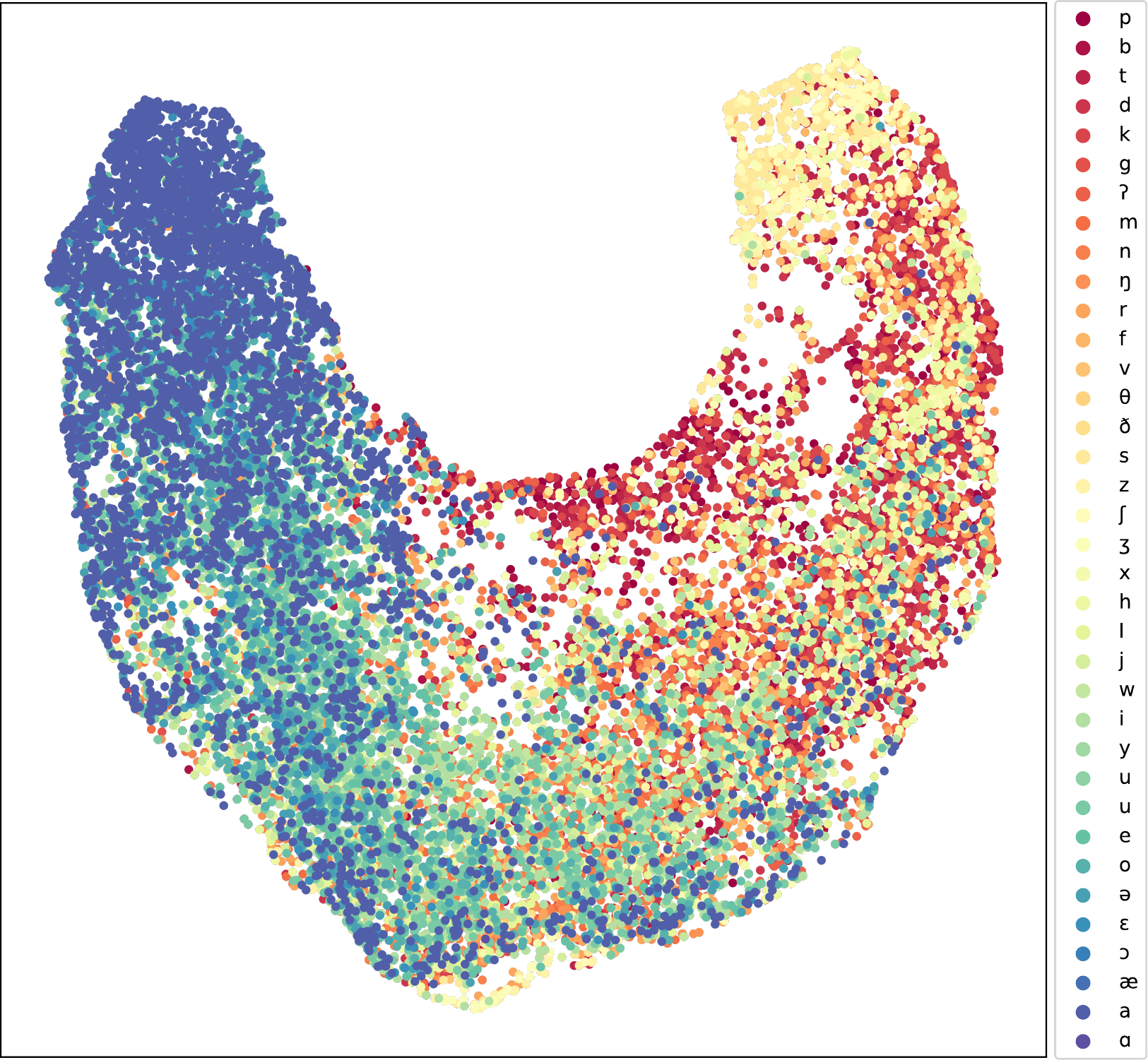}
\caption{\textbf{A Larger UMAP Projection of Encoded Vocal Tract Area Functions.} We provide a UMAP projection of an even larger set of phonemes. The embedded area functions form a continuous space spanning the set of IPA symbols. }
\label{fig:umap_big}
\end{figure}

UMAP~\citep{mcinnes2018umap-software} projections are performed on area functions encoded from the VoxAngeles dataset~\citep{Chodroff2024voxangeles}.
To generate the UMAP projections, we used 12 neighbors and a minimum distance of 0.2. However, we found that the UMAP projections are generally robust to these hyperparameters.  
In Figure~\ref{fig:umap_big}, we show an UMAP projection on an even larger set of IPA phonemes. The area functions remain disentangled by phoneme, and still form a continuous spectrum. 

\subsection{MRI Generation}
A StyleGAN2~\citep{Karras2019AnalyzingAI} network is trained on the MRI data of~\citep{Lim2021}. Using the identity labels from the dataset, the network is conditioned on these labels to preserve identity during reconstruction. to The model is trained using the default configurations for 64x64 images. During reconstruction, we use $K=30$ latent vectors to reconstruct 3 second clips at a time.

The audio recorded from the MRI dataset~\citep{Lim2021} is extremely noisy due to the spinning magnets in the MRI machine. To remove noise from the recordings, we use the Adobe Express speech enhancement tool. We then use the enhanced speech samples as the target utterances when reconstructing the MRI video. 

\subsubsection{MRI Reconstruction with Diffusion Models}
We have also tried using Diffusion Posterior Sampling (DPS)~\citep{chung2023diffusion} for MRI reconstruction, but we have found that its performance was worse than using the GAN. The diffusion model was not able to make large geometric changes such as tongue movement. DPS and its follow-ups mainly focus on low-level inverse problems such as deblurring. Compared to a diffusion model, a GAN has a semantically meaningful latent space which allows better optimization of the vocal tract shape during speech. Solving blind inverse problems with diffusion models is still an open problem, and we believe that combining diffusion with acoustic simulation is an exciting extension of our work. 

\section{Ethical Discussion}
 While the simulator presented in this paper is intended for realistic speech reconstruction and not speech generation, we acknowledge that generative speech models---and the tasks that they enable, such as voice cloning---have become serious ethical issues. As with other speech models, responsible use is essential. All samples produced with our simulator should be clearly labeled as such.

\end{document}